\documentclass[a4paper]{quantumarticle}
\pdfoutput=1

\usepackage{graphicx} 
\usepackage{xcolor}
\usepackage{listings}
\usepackage{amsmath}
\usepackage{amssymb}
\usepackage[ruled,vlined,linesnumbered]{algorithm2e}
\SetArgSty{textnormal}
\usepackage{xurl}
\usepackage{etoolbox}
\usepackage{hyperref}
\usepackage[numbers,sort&compress]{natbib}
\AtBeginEnvironment{table*}{\small\setlength{\tabcolsep}{3pt}}
\title{Entanglement-Aware Quantum Circuit Optimization through Globally Optimal Partitioning}

\author{Gregory Morse}
\orcid{0000-0002-0231-6557}
\author{Muhammad Al Farizi}
\affiliation{Department of Programming Languages and Compilers, E\"otv\"os Lor\'and University, Budapest, Hungary}

\author{Jakab Nádori}
\affiliation{Department of Physics of Complex Systems, E\"otv\"os Lor\'and University, Budapest, Hungary}
\affiliation{Qutility @ Faulhorn Labs, Budapest, Hungary}

\author{László Hajas}
\affiliation{Department of Biological Physics, E\"otv\"os Lor\'and University, Budapest, Hungary}

\author{Tam\'{a}s Kozsik}
\orcid{0000-0003-4484-9172}
\affiliation{Department of Programming Languages and Compilers, E\"otv\"os Lor\'and University, Budapest, Hungary}

\author{P\'{e}ter Rakyta}
\email{peter.rakyta@ttk.elte.hu}
\orcid{0000-0002-3506-558X}
\affiliation{Department of Physics of Complex Systems, E\"otv\"os Lor\'and University, Budapest, Hungary}
\affiliation{Computational Sciences Department, HUN-REN Wigner Research Centre for Physics, Budapest, Hungary}

\begin{document}

\begin{abstract}
    We introduce a scalable framework for globally optimal quantum-circuit partitioning and entanglement-aware block resynthesis. The partitioning problem is formulated as an integer linear program over qubit-feasible convex subcircuits, supporting objectives such as minimizing the number of partitions or a weighted execution cost. The resulting partition selection is exact over the enumerated feasible candidates and their computed costs when the ILP is solved to proven optimality. The selected partitions are resynthesized using an operator-Schmidt rank (OSR) based synthesis technique, which exploits their internal entanglement structure. We compare the end-to-end integer linear programming (ILP)--OSR workflow with BQSKit while identifying their different routing methods, and separately compare local OSR and BQSKit synthesis on matched three- and four-qubit unitaries. On the frozen sample of 772 local unitary targets, OSR produces 741 verified syntheses versus 617 for BQSKit, and matches or improves the CNOT count in all 617 jointly verified pairs. These results demonstrate that globally optimal partition selection combined with entanglement-aware resynthesis improves circuit quality and extends block-based optimization to problem sizes that are impractical for existing optimization workflows.
\end{abstract}



\maketitle

\section{Introduction}


Partitioning quantum circuits into valid multi-qubit blocks is a core primitive in modern compilation workflows or quantum computer simulation performance. It directly affects gate fusion (i.e. merging multiple gates as a single entity to reduce the arithmetic operations), and quantum circuit resynthesis with hardware-aware transpilation.
The strategy to select the partitions determines both the optimization opportunities accessible during the circuit optimization and the computational cost of exploiting them. In the noisy quantum computer era, where gate count, depth, and noise budgets are all constrained, better partitioning can translate into substantially better circuit executables.

This importance has become even clearer as quantum compilers increasingly rely on block-level numerical optimization. Parameterized circuit instantiation has emerged as a powerful primitive for both circuit optimization and gate-set transpilation, enabling global improvements beyond purely local rewrite rules \cite{Younis2022PCI}. Likewise, synthesis frameworks such as QFAST show that expressive multi-qubit building blocks combined with numerical optimization can substantially improve circuit quality while remaining scalable in practice \cite{Younis2021QFAST}. Subsequent work has pushed these synthesis perspectives in several complementary directions: LEAP improves the scalability of numerical-optimization-based synthesis through incremental prefixes and resynthesis \cite{Smith2023LEAP}, QSeed uses machine learning to seed synthesis and reduce search cost while preserving solution quality \cite{Weiden2023QSeed}, QFactor accelerates circuit instantiation itself through a domain-specific tensor-network-based optimizer \cite{Kukliansky2023QFactor}, and permutation-aware synthesis shows that jointly reasoning about block implementations and qubit permutations can improve mapping quality on constrained hardware \cite{Liu2023PAS}. Earlier block-by-block optimization frameworks such as QGo also demonstrated that partitioning can be used to scale synthesis-based optimization to circuits with tens of qubits by regenerating and recomposing independently optimized subcircuits \cite{Wu2020QGo}. For wide circuits, TopAS further demonstrates that partitioned topology-aware synthesis can reduce depth and two-qubit gate count before or alongside mapping on sparse hardware topologies \cite{Weiden2022TopAS}. These developments strengthen the case for high-quality partitioning: if downstream tools operate on blocks, then the procedure that forms those blocks and processes them becomes a decisive part of the overall compiler stack.

Several widely used frameworks already include built-in partitioning routines. Qiskit provides the \texttt{CollectMulti\allowbreak QBlocks} pass, while BQSKit offers multiple strategies such as Quick, Scan, Cluster, and Greedy. In practice, however, these approaches are heuristic and prioritize speed over global optimality. Recent work has therefore sought to improve partition quality without abandoning scalability. Clark et al. introduced Tree-based Directed Acyclic Graph (TDAG) partitioning and subsequently Greedy Topology-Aware Quantum Circuit Partitioning (GTQCP), both of which improve upon earlier heuristic baselines while preserving practical runtimes \cite{10.1145/3583781.3590234,Clark_2023}. Related directions include graph-based acyclic partitioning for simulation \cite{9912699}, Integer Linear programming (ILP) based allocation of subcircuits across GPUs in Atlas \cite{10.1109/SC41406.2024.00087}, graph partitioning for distributed quantum computation \cite{10.1117/12.3042502}, balanced partitioning via a two-level ILP \cite{Davarzani2022}, and ILP-driven routing and mapping methods that demonstrate the broader value of exact optimization in quantum circuit routing \cite{10.1145/3544563}. Partitioning is also central to approximation-oriented optimization: QuEST use partitioning to make approximate synthesis scalable and to trade fidelity against two-qubit gate count in a controlled way \cite{Patel2022QUEST}. In more application-specific settings, compiler structure can be specialized even further: Bassman Oftelie et al. give constant-depth constructions for certain dynamic materials simulations, illustrating how domain-specific circuit generation can dramatically reduce depth for targeted Hamiltonian families \cite{BassmanOftelie2022ConstantDepth}. Beyond gate count and depth, partition-aware compilation is increasingly being used to optimize other resources as well: circuit resizing in BQSKit leverages synthesis and instantiation to reduce qubit count, while Superstaq highlights how deeper, hardware-tailored optimization across abstraction layers can push more programs into the executable regime on real devices \cite{Niu2024Resizing,Campbell2023Superstaq}. 

{At a conceptual level, our approach changes how circuit optimization is framed. Rather than committing to a single partitioning heuristic and optimizing only the blocks it produces, we consider the full space of admissible ways to divide the input circuit into blocks up to a prescribed size, then select the combination with the most favorable resynthesis cost. Our results show that optimizing the partitioning objective globally over this candidate space is practical, offering a fundamentally different alternative to one-shot heuristic partitioning.}

{Making the global search practical also requires a synthesis engine that can evaluate its candidate blocks quickly and reliably. We address this with an operator-Schmidt-rank (OSR)-guided method~\cite{MullerHermes2018OSR,BalakrishnanSankaranarayanan2011OperatorSchmidt} for synthesizing small unitaries. OSR across a qubit cut can bound the number of CNOTs crossing that cut, but this bound alone does not identify the cut sequence that realizes the full unitary with the {fewest CNOT gates}.  Our key contribution is to use OSR not merely as a lower bound, but as constructive guidance for searching over CNOT structures. {OSR adds structural information to the search that a scalar Frobenius or Hilbert--Schmidt fitting cost alone does not provide~\cite{Rakyta2022approaching}.} To our knowledge, this is the first practical general-purpose synthesis method for small unitaries to use OSR in this constructive role. {{\color{black}In the matched local experiments, OSR-guided search yields lower median solver times than the tree-search synthesis used by the BQSKit baseline, supporting its use within the global circuit optimization strategy described here.}}}

This paper therefore develops {two core components}. In the first part, we introduce an OSR-based tree-search algorithm for small-partition decomposition. The method uses operator-Schmidt-rank information across subsystem cuts~\cite{MullerHermes2018OSR,BalakrishnanSankaranarayanan2011OperatorSchmidt} to derive dyadic rank certificates, weighted residual tail scores, and lower bounds on the additional CNOT gates needed to complete a candidate structure. These quantities guide branch prioritization and child generation, allowing the search to focus on CNOT structures that are most likely to reduce the remaining entanglement obstruction. In this way, the OSR strategy provides the fast and reliable local synthesis engine required for evaluating candidate partitions.

In the second part, we turn to the global partitioning problem. Given the ability to optimize small partitions efficiently, we present a practical and scalable exact algorithm for global partition optimization in quantum circuits. Our method contracts single-qubit sub-chains, enumerates all feasible convex partitions, and selects an optimal cover of the original circuit through an ILP formulation. 
The resulting approach establishes, to our knowledge, the first practical ground truth for globally optimal partitioning on circuits with up to $10^5$ gates, thereby enabling rigorous benchmarking and more principled compiler design. {Optimality refers to the enumerated blocks and their computed costs, and requires a proven optimal ILP solution.}

Both the OSR-based synthesis algorithm and the ILP-based global partitioning algorithm are implemented in the open-source SQUANDER package~\cite{SQUANDERRepository}.
Then, in the third part we summarize the results of our numerical experiments on quantum circuit optimization.

\section{OSR-guided gate synthesis strategy for topology-constrained decomposition}
\label{sec:osr-search}

For topology-constrained synthesis, where the two-qubit gate connectivity between the qubits is limited, we use an OSR-guided search over ordered, topology-respecting CNOT gate structures. In the benchmark configuration, a Hilbert--Schmidt cost function~\cite{Rakyta2022approaching} is used to optimize the single-qubit parameters of each trial gate structure. Then, operator-Schmidt spectra are evaluated to guide the discrete search over the viable CNOT gate structures decomposing the given quantum circuit. To fix notation used in this work, we recall that OSR is defined with respect to a bipartite partition of the underlying system. Suppose we have an operator
\[
A : \mathcal{H}_A \otimes \mathcal{H}_B
\rightarrow
\mathcal{H}_A \otimes \mathcal{H}_B ,
\]
where $\mathcal{H}_A$ and $\mathcal{H}_B$ denote two subsystems. The operator can be expressed in its operator-Schmidt decomposition as
\[
A = \sum_{i=1}^{r} s_i \, X_{j_i} \otimes Y_{k_i} ,
\]
where $X_i$ and $Y_i$ are operators acting on subsystems $\mathcal{H}_A$ and $\mathcal{H}_B$, respectively, and $s_i > 0$ are the operator-Schmidt coefficients.
The operator sets $\{X_j\}$ and $\{Y_k\}$ form orthonormal bases within the subspaces.
The number of nonzero coefficients $r$ is called the \emph{operator-Schmidt rank}. It quantifies the degree of coupling between the two subsystems and generalizes the concept of Schmidt rank for quantum states.
An operator with OSR equal to one can be written as a simple tensor product
\[
A = X \otimes Y ,
\]
indicating a fully separable operation. Larger operator-Schmidt ranks correspond to increasingly complex and entangling interactions between the subsystems.
In practice, the operator-Schmidt rank can be obtained by reshaping the operator tensor into a matrix representation with grouped subsystem indices and performing a singular value decomposition (SVD). The number of nonzero singular values determines the OSR.
The operator-Schmidt rank is widely used in quantum information theory, tensor-network methods, matrix product operators, and the complexity analysis of quantum circuits and many-body systems \cite{BalakrishnanSankaranarayanan2011OperatorSchmidt,MullerHermes2018OSR,Parker2020LocalMPO,Hubig2017GenericMPO,Bertini2020OperatorEntanglement}. Recent work develops a determinantal distance geometry for two-qubit synthesis~\cite{BaladaGaggioli2026LowRank}.
In this work we develop a generic N-qubit approach to guide constructive structure search for quantum gate synthesis.

The relevance of OSR to CNOT counting is well established in the two-qubit and bipartite-unitary literature. A generic two-qubit unitary requires three CNOT gates, and three are also sufficient, giving a tight non-heuristic CNOT benchmark for this smallest nontrivial case \cite{VidalDawson2004ThreeCNOT,ShendeMarkovBullock2004MinimalCNOT}. At the same time, all bipartite unitaries with operator-Schmidt rank two are locally equivalent to controlled unitaries, explaining why a single CNOT crossing a bipartition has OSR value of two and why a single cross-cut CNOT can increase OSR across that cut by at most a factor of two \cite{CohenYu2013RankTwoControlled,BalakrishnanSankaranarayanan2011OperatorSchmidt}. More generally, OSR is submultiplicative under operator composition, so $\lceil\log_2(\mathrm{OSR})\rceil$ gives a natural cut-based lower bound on the number of crossing CNOT gates needed across a partition. This perspective is consistent with standard circuit-synthesis lower-bound arguments \cite{ShendeBullockMarkov2006Synthesis} and with bipartite-unitary constructions whose entanglement or nonlocal-gate costs scale with Schmidt rank \cite{ChenYu2016EntanglementCost}.

Let $Q=\{0,\dots,n-1\}$ be the qubit set and $E_T$ the available two-qubit couplings.
We define a trial gate structure synthesizing a given circuit partition by an ordered CNOT gate structure defined by the successive two-qubit coupling edges
\begin{equation}
  P=(e_1,e_2,\dots,e_{\ell}), \qquad e_i \in E_T\;.
\end{equation}
 Elements in $P$ specify an ordered structure of CNOT gates forming a parametric circuit when surrounded with single qubit rotation gates. The circuit synthesis is performed via continuous optimization of the single-qubit gate parameters. For the results reported in this work, the continuous optimization stage uses Hilbert--Schmidt cost function. Thus, the method of gate synthesis alternates between a discrete combinatorial stage (the CNOT gate structure) and a continuous parameter tuning process.

For the fitted trial circuit $V_P$ (i.e. approximating the effect of the unitary $U$ with the tiral $V_P$ circuit), the fidelity of the approximation is evaluated by Hilbert--Schmidt cost funtion
\begin{equation}
 h(P)=1-\frac{|\operatorname{Tr}(V_PU^\dagger)|^2}{4^n},
\end{equation}
with $n$ being the number of qubits.
The cut profiles defined below are evaluated on the residual $V_PU^\dagger$. 

Let $C\subset Q$ be a nontrivial bipartition of the qubit register $Q$ (made of the $n$ qubits) and let $\overline{C} = Q\setminus C$, being its complementary. 
The first step is to permute the rows and columns of the unitary $U$ to separate the partitions $C$ and $\overline{C}$. The permutation according to the subset $C$ is described by the operator $\mathcal{R_C}$:
\begin{equation}
\label{eq:osr-reshape}
  \mathcal R_C(U)_{(a',a),(b',b)} = U_{(a',b'),(a,b)},
\end{equation}
where $(a,a')\in \{0,\dots,d_C-1\}^2$ index the $C$ subsystem and $(b,b')\in \{0,\dots,d_{\overline{C}}-1\}^2$ index the complement. 
Finally, we compute a standard thin SVD,
\begin{equation}
  R_C(U) = \widetilde U_C\,\Sigma_C\,\widetilde V_C^{\dagger}\;,
\end{equation}
with singular values
\begin{equation}
  \begin{aligned}
  \sigma_0^{(C)} &\ge \sigma_1^{(C)} \ge \dots \ge \sigma_{k_C-1}^{(C)} \ge 0,\\
  k_C &= \min(d_C^2,d_{\overline{C}}^2)\;,
  \end{aligned}
\end{equation}
where $d_C=2^{|C|}$ and $d_{\overline{C}}=2^{|\overline{C}|}$ are the dimensions of the Hilbert-space associated with the cuts $C$ and $\overline{C}$.
(Thin SVD keeps only the useful components from the singular value decomposition of $R_C(U)$)
In our implementation in the SQUANDER package these singular values are normalized by the Frobenius norm of the target unitary,
\begin{equation}
  \widehat \sigma_j^{(C)} = \frac{\sigma_j^{(C)}}{\|U\|_F},
  \qquad \|U\|_F \equiv \sqrt{2^n}\;,
\end{equation}
so all OSR-derived quantities are comparable across different cuts.

\subsection{OSR-Based Tree-Search Strategy with Branch Prioritization}

To describe our circuit synthesis strategy, we first introduce the following quantities.

\paragraph{Discrete dyadic rank certificate:} After solving the continuous optimization problem of the single-qubit rotation angles across a trial CNOT structure, we examine the resulting residual unitary $V_PU^\dagger$ across all nontrivial qubit-register cuts. Let $\mathcal C$ denote the set of nontrivial unordered bipartitions. For each cut $C\in\mathcal C$, we compute the normalized singular values $\widehat \sigma_j^{(C)}$ using the SVD procedure described in Sec.~\ref{sec:osr-search} and define the numerical OSR count as
\begin{equation}
  r_C^{\mathrm{num}}(U)
  = \#\bigl\{j : \widehat \sigma_j^{(C)} > \widehat \sigma_0^{(C)}\,\tau\bigr\}.
\end{equation} 
$r_C^{\mathrm{num}}$ evaluates the number of singular values for cut $C$ that are larger than $\tau$-times the largest singular value, with $0\leq\tau<1$.
Based upon this quantity, we introduce the so-called \emph{discrete dyadic certificate} used in the tree search, which is calculated as
\begin{equation}
\label{eq:dyadic-rank-bound}
  b_C(U) = \left\lceil \log_2\bigl(r_C^{\mathrm{num}}(U)\bigr) \right\rceil.
\end{equation}
This is the quantity returned as the OSR-based lower bound on the number of crossing CNOT gates required to disentangle the fitted residual across cut $C$\cite{CohenYu2013RankTwoControlled,BalakrishnanSankaranarayanan2011OperatorSchmidt}.

\paragraph{Weighted Tail Loss:} 
{The weighted tail loss provides a secondary score for ranking trial gate structures:}
\begin{equation}
\label{eq:kappa-def}
 \kappa_C(U)=\sum_{j=2^{\beta_C(U)}}^{k_C-1}
 \rho^{\,k_C-1-j}\bigl(\widehat{\sigma}_j^{(C)}\bigr)^2,
 \qquad \rho=0.1,
\end{equation}
with
\begin{equation}
 \beta_C(U)=
 \begin{cases}
 b_C(U),&n\leq 3,\\
 \max\{b_C(U)-1,0\},&n\geq 4.
 \end{cases}
\end{equation}

The largest tail weights belong to the smallest singular values. {The weighted tail score distinguishes trial gate structures with the same OSR lower bound and helps determine which structures the search explores first. It is a heuristic ordering score and does not give the exact number of CNOT gates needed to complete the decomposition. Its role in choosing additional CNOT connections is described below.}
For a given trial CNOT gate structure $P$ and optimized rotation parameters indexed by $C_0$, we introduce the full cut profile
\begin{equation}
  \Gamma_{C_0}(P)=\bigl\{\bigl(b_C(P;{C_0}),\kappa_C(P;{C_0})\bigr): C\in\mathcal C\bigr\}\;.
\end{equation}
This profile is what drives the discrete tree-search over the possible gate structures. The Hilbert--Schmidt fit improves a fixed structure, while the all-cut OSR post-evaluation estimates an optimistic lower bound on the additional CNOT gates required to complete it.

\paragraph{Estimation on the number of additional CNOT gates required to finish the synthesis:}
The cut profile is used to determine a combinatorial lower bound on the crossing interactions needed to synthesize the current fitted residual. For any cut $C\in\mathcal C$, let
\begin{equation}
  \delta_T(C)=\bigl\{\{u,v\}\in E_T : u\in C,\; v\notin C\bigr\}
\end{equation}
be the set of two-qubit connection edges crossing $C$. If a future modification of the trial circuit adds $x_e\in\mathbb{Z}_{+,0}$ CNOTs on edge $e\in E_T$, then the induced cut coverage is
\begin{equation}
  \operatorname{cov}_{\mathbf{x}}(C)=\sum_{e\in\delta_T(C)} x_e.
\end{equation}
Since the dyadic rank certificate gives an optimistic lower bound on the number of CNOTs that must cross each cut, a circuit realizing that fixed residual must satisfy
\begin{equation}
  \operatorname{cov}_{\mathbf{x}}(C)\ge b_C(P;C_0), \qquad \forall C\in\mathcal C.
\end{equation}
Thus the lower bound on the required additional CNOT count is obtained from the optimization problem
\begin{equation}
\label{eq:osr-min-cnot-bound}
  L_{C_0}(P)= \min_{\mathbf{x}\in\mathbb{Z}_{+,0}^{|E_T|}}
  \sum_{e\in E_T} x_e
\end{equation}
subject to
\begin{equation}
  \sum_{e\in\delta_T(C)} x_e \ge b_C(P;C_0),\qquad \forall C\in\mathcal C. \label{eq:osr-min-cnot_constraint}
\end{equation}
This lower-bounding subproblem is solved by an exact combinatorial search over edge multiplicities $x_e$. It relaxes ordering and realizability constraints for the measured residual. Jointly refitting a child can change that residual, so this value is structural guidance rather than an admissible pruning certificate for every refitted descendant. The hard CNOT-budget check uses the actual path length, not an assumed admissible bound for all refitted descendants.

\paragraph{Secondary ordering via the weighted tail loss:} The minimum in~\eqref{eq:osr-min-cnot-bound} is typically highly degenerate. To break ties meaningfully, the implementation solves a second lexicographic problem after the minimum CNOT count has been fixed. For a feasible edge-count vector $\mathbf{x}=\{x_{e_1}, x_{e_2}, \dots \}$ with $\sum_e x_e=L_{C_0}(P)$, define the surplus on cut $C$ by
\begin{equation}
  \operatorname{sur}_{\mathbf{x}}(C)=\operatorname{cov}_{\mathbf{x}}(C)-b_C(P;C_0)\ge 0.
\end{equation}
Among all minimizers of~\eqref{eq:osr-min-cnot-bound} we then minimize
\begin{equation}
\label{eq:osr-surplus}
  \begin{aligned}
    S_{C_0}(P)=\min_{\mathbf{x}} \quad & \sum_{C\in\mathcal C} \kappa_C(P;C_0)\,\operatorname{sur}_{\mathbf{x}}(C) \\
    \text{subject to} \quad & \sum_{e\in E_T}x_e=L_{C_0}(P), \\
    & \operatorname{cov}_{\mathbf{x}}(C)\ge b_C(P;C_0), \qquad \forall C\in\mathcal C.
  \end{aligned}
\end{equation}
This step is where the weighted tail loss $\kappa$ becomes critical. Cuts with the same dyadic bound $b_C$ need to be classified differently: a cut whose spectrum still carries a large weighted tail should influence the search more strongly than one whose post-threshold tail is already nearly extinguished. Equation~\eqref{eq:osr-surplus} encodes exactly this preference. The optimization result $\mathbf{x}_{C_0}^{\star}(P)$ (for which $S_{C_0}(P)$ is minimal) is then reused to rank topology edges (translated to new CNOT gate candidates) {when adding CNOT gates to the trial structure}.

\paragraph{Ordering the trial gate structures: } 
Each fit $C_0$ yields a tuple
\begin{equation}
  \Xi_{C_0}(P)=\bigl(L_{C_0}(P),\,S_{C_0}(P),\,x_{C_0}^{\star}(P),\,\Gamma_{C_0}(P)\bigr).  
\end{equation}
This tuple records, for one continuous fit, the optimistic residual CNOT bound $L_{C_0}(P)$, the weighted surplus tie-breaker $S_{C_0}(P)$, an optimal edge-count assignment $x_{C_0}^{\star}(P)$ for prioritizing topology edges, and its measured all-cut profile $\Gamma_{C_0}(P)$. 
(In the procedure we may run the continuous optimization multiple times and store only the best result.)

For repeated fits of one structure, the best profile is selected lexicographically:
\begin{equation}
 K_{\mathrm{fit}}(P)=
 \min_{C_0}\bigl(L_{C_0}(P),S_{C_0}(P),\Phi_{C_0}(P)\bigr),
\end{equation}
where
\begin{equation}
\label{eq:phi-tertiary-key}
 \Phi_{C_0}(P)=\sum_{C\in\mathcal C}
 \left(|\mathcal C|\,b_C(P;C_0)+\kappa_C(P;C_0)\right).
\end{equation}
Write $(\bar L,\bar S,\bar\Phi)=K_{\mathrm{fit}}(P)$. The actual comparison keys between different structures are
\begin{equation}
 K_n(P)=
 \begin{cases}
 (|P|+\bar L,\bar L,\bar S,\bar\Phi),&n=3,\\
 (\bar L,h(P),\bar S,|P|),&n=4.
 \end{cases}
\end{equation}
Thus both widths use the same fitted-residual and cut-cover framework, but the four-qubit policy places the fitted HS score immediately after the residual CNOT bound. These are ordering rules inside an optimistic forward search with deferred candidates, not a claim of exhaustive A$^\ast$ optimality or backtracking. The external CNOT budget remains distinct from the residual estimate.

\paragraph{Heuristic to extend trial gate structures: } After sorting and ordering the already visited gate structures, now we discuss the developed heuristic approach to further evolve the most promising gate structures (i.e. adding new CNOT layers to the gate structure). 
Once the current best CNOT gate structure $P$ has been identified during the gate synthesis, the search expands it by inserting additional CNOTs into the current ordered CNOT gate structure. The expansion mechanism is guided by the edge-count assignment vector $x_A^{\star}(P)$ associated with the chosen best gate structure, as the anchor of further extensions.

The fitted cut profile and its edge-cover assignment rank legal extensions. Single-CNOT additions are considered before more structured multi-CNOT insertion classes; constructive pair and motif proposals supplement this ordering. A deferred frontier avoids fitting every possible placement before trying a promising extension. These priorities are heuristic, rather than an exhaustive ordering of every insertion by one edge-demand vector.

Formally, for a chosen CNOT insertion multiplicity $m$, the neighborhood explored is a constrained subset
\begin{equation}
  \begin{aligned}
  \mathcal{N}_m\bigl(P\,;x_{C_0}^{\star}(P)\bigr)
  &\subseteq \bigl\{\text{all ways to insert $m$ CNOTs}\\
  &\qquad\text{into gate structure } P\bigr\},
  \end{aligned}
\end{equation}
where the ordinary insertion generator uses the following guiding principles:

\begin{itemize}
 \item[G1] \emph{Spatial coherence of inserted edges:} After the first CNOT insertion, every subsequent inserted CNOT must share at least one qubit with a CNOT already inserted in the same expansion call. In other words, the set of newly inserted CNOTs is required to form a connected pattern in the line graph of the topology. This biases the search toward coherent local entangling motifs instead of disjoint edits.
 \item[G2] \emph{Repetition cap:} Any candidate path containing four consecutive identical (or reverse-direction) CNOTs is discarded immediately, since any two-qubit unitary can be synthesized with at most three CNOT gates~\cite{VidalDawson2004ThreeCNOT,VatanWilliams2004OptimalTwoQubit,ShendeMarkovBullock2004MinimalCNOT}. Pruning these candidates early removes a large family of syntactically distinct but practically useless trial gate structures.
 \item[G3] \emph{Canonical uniqueness filtering:} Before evaluation, each candidate is passed through a structural uniqueness test. Ordered CNOT structures that are syntactically different but represent the same effective two-qubit interaction pattern are rejected, and a local visited set suppresses duplicates generated through different insertion orders. This canonicalization is essential for keeping the search tree from filling with isomorphic gate structures.
\end{itemize}
These guiding principles do not affect the validity of the lower bound in (\ref{eq:dyadic-rank-bound}); they only restrict which children are materialized. Hence the combinatorial explosion of naive insertion search is kept under control.

\subsection{Summary of the OSR based search procedure}

\begin{algorithm}[t]

\DontPrintSemicolon
\caption{OSR-guided synthesis with certified constructive completion}
\label{alg:osr-best-first}
\SetKwInOut{KwIn}{Input}\SetKwInOut{KwOut}{Output}
\KwIn{Target unitary $U$; topology $T=(Q,E_T)$; CNOT budget $L_{\max}$.}
\KwOut{A certified topology-respecting decomposition, or failure within the budget.}
$\mathcal C\gets$ all nontrivial qubit cuts\;
{Try the complete trial gate structures described above, using at most $L_{\max}$ CNOTs}\;
\If{a completion is certified}{
 \Return the completion, {after testing CNOT removals and refitting when applicable}\;
}
$P\gets\emptyset$; fit its parameters and measure its all-cut residual profile\;
\If{$P$ passes full-unitary certification}{\Return $P$\;}
\While{an untried legal forward extension within $L_{\max}$ remains}{
 Generate or resume the next deferred structured insertion class anchored at $P$\;
 Fit the next candidate with bounded HS optimization and evaluate its all-cut profile\;
 \If{the candidate passes full-unitary certification}{
  \Return the candidate, {after testing CNOT removals and refitting when applicable}\;
 }
 Rank the candidate using $K_n$ and the structured class priorities\;
 \If{the forward-progress conditions are met}{
  Commit the candidate as the new current path $P$\;
 }
 Otherwise continue the deferred class or the next insertion multiplicity\;
}
\Return failure\;

\end{algorithm}

The resulting control flow is summarized in Algorithm~\ref{alg:osr-best-first}.
A zero cut-cover bound alone does not prove that the decomposition is valid. A candidate is accepted only if a fixed-structure Hilbert--Schmidt fit reaches the synthesis tolerance. Otherwise, the search continues until the CNOT budget is exhausted. For both three- and four-qubit partitions, the search uses post-fit OSR bounds, $\kappa$-weighted ordering, and structured child generation, with the width-dependent keys above and the same strict certification requirement. It does not guarantee a minimum-CNOT solution.

\paragraph*{Complexity considerations.}
The dominant cost of one node evaluation is the repeated BFGS refinement together with the SVDs needed to evaluate the induced all-cut profile. By contrast, the combinatorial solver for~\eqref{eq:osr-min-cnot-bound} is inexpensive because the topology size is fixed and small relative to the ambient unitary dimension. The search is therefore designed to spend substantial floating-point effort only on nodes that survive the strong discrete guidance. This is precisely why the definition of $\kappa$ in~\eqref{eq:kappa-def} and the $\kappa$-weighted surplus objective~\eqref{eq:osr-surplus} are not cosmetic details: they are what turn a coarse dyadic OSR certificate into a practically discriminative best-first ordering.

\section{From OSR-Guided Local Synthesis to Global Circuit Optimization via Partitioning}

The OSR-guided search described above provides the local synthesis engine used to evaluate small subcircuits. Its purpose is to decide, for a fixed set of qubits and a fixed hardware topology, which CNOT structures are promising enough to be refined by continuous single-qubit gate parameter optimization. The global problem considered in the remainder of the paper is complementary: instead of asking how to decompose one small unitary, we ask how a large circuit DAG should be divided into subcircuits so that each subcircuit can be optimized locally while the collection of optimized subcircuits will lead to the most compressed quantum circuit. This section introduces the graph-theoretic and optimization concepts that connect these two levels.

To formalize the global step, we model the input circuit as a directed acyclic graph (DAG)\cite{vandegriend2024comparisonquantumcompilersusing,childs_et_al:LIPIcs.TQC.2019.3}, whose vertices are gates and whose edges encode gate dependencies. A candidate partition is useful only if it can be treated as a self-contained subcircuit to be passed to the local synthesis procedure. This requirement is captured by closed convexity in the DAG: whenever two gates belong to the same partition, every directed path between them must remain inside the partition. Thus, in our formulation, valid circuit partitions correspond to closed convex sets of the DAG. Open convex sets can also be enumerated efficiently \cite{BALISTER2009509}, but they are not appropriate here because they may split a partition into disconnected fragments that no longer define a single coherent local synthesis task.

The OSR-based decomposition routine is intended for small partitions, so every candidate partition is additionally restricted by a maximum number of qubits. This bound is both algorithmic and physical: it keeps the OSR-guided synthesis problem small enough to solve reliably, and it reflects hardware and simulation constraints on the number of qubits that should be optimized as one block. In practice, this limit is usually in the range of 3--5 qubits. The special case of two-qubit blocks is comparatively simple, as any two-qubit unitary can be decomposed at most $3$ CNOT gates\cite{tucci2005introductioncartanskakdecomposition}.

Efficient enumeration and validation of the partitions relies on standard graph operations. The transitive closure of a DAG introduces an edge $(u,v)$ whenever a path $u\rightarrow v$ exists in the original graph.
This makes convexity and dependency checks easier to express. We compute the closure information using Nuutila's transitive-closure approach~\cite{NUUTILA1994207}. Its strongly connected component (SCC) subroutine admits linear-time implementations, such as Tarjan's algorithm~\cite{doi:10.1137/0201010} and textbook variants~\cite{aho1974design}.

Once all feasible convex partitions of the original circuit have been generated, the remaining task is to select the subset that yields the optimally compressed quantum circuit.
We formulate this global optimization step as an Integer Linear Programming (ILP) problem \cite{doi:https://doi.org/10.1002/9781118627372.ch1}.
All candidate subcircuits, including overlapping and non-overlapping ones, are enumerated subject to the maximum qubit budget.
Each candidate subcircuit is then optimized using the OSR-based decomposition routine, and the resulting costs are used to determine an optimal exact cover of the original circuit.
In other words, to reproduce the original quantum program, we select a subset of optimized subcircuits that minimizes the total number of CNOT gates.
The ILP constraints ensure that every gate in the original circuit is assigned to exactly one selected partition, while the objective function encodes the desired global cost, such as the CNOT count of the resulting circuit.



For completeness, additional preprocessing and postprocessing routines are provided in the appendices. Appendix~\ref{apx:contract} details the contraction of single-qubit chains, while Appendix~\ref{apx:recomb} explains their recombination after ILP optimization. In brief, single-qubit chains without parents or children form partitions trivially. Chains with only one neighbor are joined to that neighbor’s partition, whereas chains with both a parent and a child require a choice. In our implementation we consider both of the choices equivalently good. 

\subsection{Identifying convex partitions in the circuit}

When exploring feasible partitions in the circuit, it is essential to enforce constraints such as the maximum number of qubits $Q_{\max}$ allowed in a single partition.
A purely structural reachability analysis may admit candidate partitions that exceed this qubit budget. To avoid such candidates, we use a procedure that propagates only through subgraphs of the DAG whose cumulative qubit sets remain within the prescribed limit.

Formally, let $G = (V,E)$ denote the gate-dependency DAG, where $V$ is the set of gates in the circuit to be optimized and $E$ is the set of dependency edges between them. Let $g(v)$ and $rg(v)$ denote the forward and reverse adjacency sets of a gate $v$, respectively.
Each gate $v \in V$ is associated with the set of qubits $\mathcal{Q}(v) \subseteq \mathcal{Q}$ on which it acts.
The procedure explores backward reachability under the constraint
\begin{equation}
  \left|\bigcup_{u \preceq v} \mathcal{Q}(u)\right| \;\leq\; Q_{\max},
\end{equation}
where $u \preceq v$ indicates that $u$ is on a valid backward path from $v$.

Algorithm~\ref{alg:qubit_reach}, given in Appendix~\ref{apx:qubit-number-aware-enum}, implements the outlined reachability procedure, ensuring that only gates belonging to subcircuits whose total qubit footprint does not exceed $Q_{\max}$ are considered reachable. Thus, infeasible sub-circuits are pruned early, substantially reducing the search space compared to unrestricted reachability.

The identified partitions are also filtered according to their convexity.
A partition $P \subseteq V$ is convex if, whenever $u,v \in P$ and $u \preceq w \preceq v$, then $w \in P$ as well. 
Convexity guarantees that partitions correspond to contiguous subgraphs without ``holes,'' while qubit number-awareness ensures that no partition exceeds the imposed bound $Q_{\max}$.
We improve the efficiency of the convex set enumeration algorithm \cite{BALISTER2009509} by maintaining the sets of forward-reachable vertices, denoted by $A$, and backward-reachable vertices, denoted by $B$.
We also track $Q$, the current set of qubits in the candidate partition.
The current partition is then greedily extended by appending gates from the frontier sets $A$ and $B$.
The complete algorithm with a pseudocode for this enumeration procedure is given in Algorithm~\ref{alg:partition_enum} in Appendix~\ref{apx:qubit-number-aware-enum}.

\subsection{Formulating the optimization problem to find the optimal partitioning}\label{sec:ilp}

The previous sections described how to generate all feasible subcircuit partitions of a quantum circuit and how the OSR-guided strategy recompiles them.
We now formulate an ILP optimization problem that selects a subset of these optimized subcircuits whose recomposition preserves the original quantum program while optimizing a chosen global metric, such as the overall CNOT gate count used in this work.
Let $V$ denote the set of gates in the original circuit, and let $\mathcal{P} = \{P_1, P_2, \ldots, P_m\}$ denote the set of all candidate unoptimized partitions.
We write $\widetilde{\mathcal{P}}$ for the corresponding set of optimized partitions, where $\widetilde{P}_i \in \widetilde{\mathcal{P}}$ is the resynthesized version of the subcircuit $P_i \in \mathcal{P}$.
Each partition $P_i$ contains a subset of the gates in $V$.
To reconstruct a circuit equivalent to the original quantum program, the selected optimized partitions must correspond to unoptimized partitions that do not share gates.
To achieve this, lets first construct the sets $\mathcal{P}(g)$ for all $g\in V$ describing which partitions $P_i\in\mathcal{P}$ contain the gate $g$.
Thus, for every gate $g \in V$, let 
\begin{equation}
  \mathcal{J}(g) = \{ j \mid g \in P_j \}  
\end{equation}
be the set of partition indices $j$ for which $g\in P_j$.
For each partition $P_j \in \mathcal{P}$ we introduce a binary variable $x_j = 1$ if partition $P_j$ is selected for the circuit reconstruction and $0$ otherwise.
Then the ILP optimization is formulated as
\begin{equation}
\min\left(\sum_{j=1}^{|\mathcal{P}|} x_j\cdot \mathcal{CNOT}(\widetilde{P}_j)\right) \label{eq:obj} 
\end{equation}
subjected to 
\begin{equation}
\sum_{j \in \mathcal{J}(g)} x_j = 1, 
\qquad \forall g \in V, \label{eq:cover} 
\end{equation}
Constraint~\eqref{eq:cover} ensures that each gate is covered by exactly one selected partition in the recombination of the circuit.
The objective~\eqref{eq:obj} minimizes the total optimized CNOT count, where $\mathcal{CNOT}(\widetilde{P}_j)$ denotes the number of CNOT gates in the optimized partition $\widetilde{P}_j$.
This objective can be replaced by any suitable cost function that reflects the goal of the circuit optimization.
For example, one may minimize the count of another hardware-expensive gate or use a cost function that models the cumulative noise of the gates in the reconstructed circuit.
For simplicity, in this work we focus on the CNOT gate count.

\subsection{Enforcing acyclicity among partitions}
\label{sec:acyclicity}

\begin{figure}[h]
  \centering
  \includegraphics[width=0.25\textwidth]{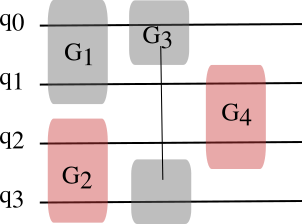}
  \caption{An example showing 2-long cycle between convex partitions. 
  Gate $G_3$ in the black partition depends on the execution of gate $G_2$ in the red partition, while gate $G_4$ in the red partition depends on the execution of gate $G_1$ in the black partition.}
  \label{fig:cyclic_example}  
\end{figure}

The procedure described above minimizes the selected cost over the enumerated, resynthesized partitions, but it can also introduce inconsistencies in the ordering of gate operations.
If the selected partitions cannot be arranged in a valid gate-execution order, the resulting circuit may implement a different quantum program from the original circuit.
Figure \ref{fig:cyclic_example}. shows an example of such ill-conditioned case.
The figure shows two partitions, highlighted with black and red colors.
Although both partitions are mathematically convex, they induce a cyclic dependency etween each other.
Specifically, gate $G_3$ in the black partition depends on the execution of gate $G_2$ in the red partition, while gate $G_4$ in the red partition depends on the execution of gate $G_1$ in the black partition.
Thus, the red partition depends on the black partition and vice versa, so the two partitions cannot be ordered acyclically.
Generally speaking, even if each partition is individually convex and locally optimizable by the OSR-based synthesis procedure, the resulting graph induced by the selected optimized partitions must originate from partitions being in acyclic relation to preserve a valid execution order. 
Cycle-inducing selections can become more frequent as the maximum qubit limit per partition grows, so the algorithm must detect and eliminate these cases. 
In our implementation provided within the SQUANDER package, we use cycle-elimination ideas related to chordless-cycle enumeration algorithms such as those of Uno and Satoh \cite{10.1007/978-3-319-11812-3_27} and Dias et al. \cite{dias2014efficientenumerationchordlesscycles}, which are also reflected in practical graph libraries such as \texttt{NetworkX}.

This section formalizes the concept of the partition--interaction digraph and presents three complementary strategies to enforce acyclicity on the solution of the ILP problem.

\paragraph{\textbf{Strategy I: Static cycle-elimination constraints.}}
Let $G=(V,E)$ be again the the original DAG and let $\mathcal{P}=\{S_1,\dots,S_m\}$ be the set of candidate partitions for the ILP problem. 
Lets define the \emph{partition--interaction directed graph} as
\begin{equation}
  H = ( \{1,\dots,|\mathcal{P}|\}, A),
\end{equation}
where $\{1,\dots,|\mathcal{P}|\}$ describes the set of the partitions through their indices from $\mathcal{P}$ and $A$ describes the directed arcs between the elements.
An arc $(i,j)$ is element of $A$ if and only if there exist edges $(u,v)\in E$ such that
\begin{equation}
  u \in S_i \setminus S_j,\qquad v \in S_j \setminus S_i,\qquad i\neq j.
\end{equation}
That is, we add an arc whenever the original graph contains a cross edge from the exclusive part of $S_i$ to the exclusive part of $S_j$ (with $S_i$ and $S_j$ being two subcircuits.).
This excludes edges attributable to partition overlap, which are already handled by the base ILP through the explicit non-overlap constraint.
In the implementation, we further prune $H$ by computing its strongly connected components (SCCs) and retaining only arcs internal to those components, since arcs crossing distinct SCCs cannot participate in any directed cycle.
(An SCC is a maximal set of vertices in which every vertex can reach every other vertex via directed paths. Therefore, vertices outside SCCs cannot belong to the same directed cycle.)
Additionally, we use the exclusive-part rule (\ref{eq:cover}) above to avoid spurious arcs due to overlaps. 
We can drop arcs between pairs that are already mutually forbidden by the base ILP (e.g., when $S_i\cap S_j\neq\emptyset$ and $x_i+x_j\le 1$ is present).

As before in  Sec.~\ref{sec:ilp}, let $x_i\in\{0,1\}$ denote whether partition $S_i$ is selected to be used in the recombination of the resulting circuit or not.
For any directed simple cycle $C$ in the directed graph $H$, the inequality
\begin{equation}
  \sum_{i\in C} x_i \;\le\; |C| - 1
  \label{eq:cycle-cut}
\end{equation}
forbids selecting all vertices of $C$ simultaneously, thereby eliminating that cycle.
Here, $C$ is a set of indices whose corresponding partitions form a directed cycle of length $|C|$.
However, pre-enumerating all such cycles can scale exponentially in the worst case.
Therefore, when computational budget is a concern, one may restrict the enumeration to cycles up to a prescribed length. Important special cases include:
\begin{itemize}
  \item cycles of length $2$: if $(i,j)\in A$ and $(j,i)\in A$, then $x_i + x_j \le 1$;
  \item cycles of length $3$: if $(i,j),(j,k),(k,i)\in A$, then $x_i + x_j + x_k \le 2$.
\end{itemize}
When cycle enumeration is limited to a prescribed length, this approach must be combined with Strategy~II or~III, discussed below, to guarantee global acyclicity.

\paragraph{\textbf{Strategy II: Introduction of continuous order variables to the partitions.}}
The idea in this strategy is to assign new continuous variables $u_i\in[0,|\mathcal{P}|]$ to each partition and compose such constraints between these variables which cannot be fullfilled for partitions forming a cycle.
Such a constraint can be given with indicator constraints
\begin{equation}
  (x_i=1)\wedge(x_j=1)\ \Rightarrow\ u_j \ge u_i + 1,
\end{equation}
for every arc $(i,j)\in A$.
If both $(i,j)$ and $(j,i)$ belong to $A$, forming a cycle of length $2$, these constraints cannot be satisfied simultaneously: summing them yields $u_i + u_j \ge u_i + u_j + 2$.
The same argument extends to cycles of arbitrary length.
This yields a compact, single-shot model with $O(|A|)$ additional constraints. It is attractive when the pruned arc set $A$ is moderate, while for extremely large $A$ the extra constraints may still be heavy.

\paragraph{\textbf{Strategy III: Lazy cycle cuts (cutting-plane enforcement).}}
In contrast to Strategies I and II, this strategy eliminates cycles only when they are discovered in a currently best solution during optimization.
Starting from the base ILP formulation without acyclicity constraints, whenever the solver finds a new best solution $\bar x$, we consider the subgraph induced by the selected partitions,
\begin{equation}
  H[\bar X] \quad \text{with} \quad \bar X = \{\, i \mid \bar x_i = 1 \,\}.
\end{equation}
If $H[\bar X]$ contains a SCC of size at least $2$, we extract a simple directed cycle $C$ from that SCC and add the corresponding cycle-cut constraint~\eqref{eq:cycle-cut}.
We then re-optimize the problem and repeat this procedure until the currently best solution contains no cycle violations.
Modern solvers, such as Gurobi, support this approach through a \emph{lazy-constraint callback} at the \texttt{MIPSOL} event, which allows on-the-fly SCC detection and insertion of~\eqref{eq:cycle-cut} before the incumbent is accepted.

For practical considerations, we mention that if $A$ becomes modestly large after pruning, Strategy~II is simple and effective. 
For very large graphs, though, Strategy~III generally yields the best time--memory trade-off. Strategy~I is useful for adding a small library of short-cycle cuts upfront (e.g., all 2- and 3-long cycles), and can be combined with strategies II or III.
All three strategies produce solutions in which the partition--interaction digraph induced by selected partitions is acyclic, thereby satisfying the decomposition requirement.

\section{Experimental Results}

We evaluated the proposed entanglement-driven OSR decomposition using its implementation in the SQUANDER package. SQUANDER combines OSR-based partition resynthesis with the ILP based partitioning formulation described above, which selects a globally optimal circuit partitioning with respect to the chosen cost function. In these experiments, the objective is to maximize circuit compression by reducing the number of CNOT gates. To assess the impact of these newly developed techniques, we compare the resulting circuit-optimization workflow with the well-established Berkeley Quantum Synthesis Toolkit (BQSKit)~\cite{Younis2021BQSKit}.

{{\color{black}The BQSKit reference uses SeqPAM~\cite{Liu2023PAS} to map blocks produced by its heuristic partitioning stage. SQUANDER retains all enumerated admissible routing blocks and their synthesized input and output permutation variants in a shared library. ILP-based partition selection propose alternative complete circuit partitionings, with cost criteria including cached CNOT counts and estimates of qubit movement (i.e. permutations on the edges of the selected partitions). PAM-style mapping and bounded beam search then choose compatible block implementations, qubit permutations, and SWAPs. We retain the configuration with lowest CNOT count, while other target function could similarly work.} We then run a final connectivity-aware optimization and resynthesis pass to reduce routing overhead associated with the extra inserted SWAP gates. This is reported as the \emph{Final} stage in the tables. The end-to-end comparison therefore reflects the combined effects of partition selection, resynthesis, and routing, rather than isolating any one component. {\color{black}Both workflows are evaluated on a linear connectivity topology, but applicable on any arbitrary cnnection between the qubits.}}


The circuit optimization workflow consists of three stages. First, the circuit is optimized under the assumption of all-to-all qubit connectivity. This stage is denoted by \emph{A2A} {\color{black}in Tables~\ref{tab:osr-bqskit-three-qubit-ibm-eagle-counts}--\ref{tab:osr-four-qubit-unpaired-qasm-benchmarks}, which summarize our results.} In the SQUANDER workflow, this stage uses ILP-based partition selection and OSR resynthesis. Second, the resulting circuit is mapped to the linear target topology using the routing method of its respective workflow. This stage is denoted by \emph{Routed} in the tables. Third, the routed circuit undergoes a final connectivity-aware partition optimization and resynthesis. This stage is denoted by \emph{Final}. {\color{black}In the SQUANDER workflow, the final stage again uses ILP-based partition selection and OSR resynthesis.}

\begin{table*}[t]
\centering
\begin{tabular}{lrr|rrr|rrr}
\hline
Circuit & Qubits & Initial CNOTs & \multicolumn{3}{c|}{Squander CNOTs} & \multicolumn{3}{c}{BQSKit CNOTs} \\
 & & & A2A & Routed & Final & A2A & Routed & Final \\
\hline
\texttt{tof\_3} & 5 & 18 & \textbf{14} & \textbf{17} & \textbf{17} & 18 & 21 & 19 \\
\texttt{barenco\_tof\_3} & 5 & 24 & \textbf{18} & \textbf{23} & \textbf{23} & 22 & 26 & 26 \\
\texttt{mod5\_4} & 5 & 28 & \textbf{20} & \textbf{27} & \textbf{25} & 22 & 30 & 29 \\
\texttt{tof\_4} & 7 & 30 & \textbf{22} & \textbf{29} & \textbf{29} & 28 & 34 & 32 \\
\texttt{tof\_5} & 9 & 42 & \textbf{30} & \textbf{41} & \textbf{41} & 40 & 49 & 46 \\
\texttt{barenco\_tof\_4} & 7 & 48 & \textbf{34} & \textbf{46} & \textbf{46} & 42 & 52 & 48 \\
\texttt{mod\_mult\_55} & 9 & 48 & \textbf{40} & \textbf{70} & \textbf{67} & 48 & 82 & 79 \\
\texttt{QFT8} & 8 & 56 & \textbf{52} & \textbf{73} & \textbf{72} & 56 & 85 & 77 \\
\texttt{vbe\_adder\_3} & 10 & 70 & \textbf{46} & \textbf{64} & \textbf{61} & 47 & 77 & 68 \\
\texttt{barenco\_tof\_5} & 9 & 72 & \textbf{50} & \textbf{68} & \textbf{68} & 70 & 83 & 79 \\
\texttt{csla\_mux\_3\_original} & 15 & 80 & \textbf{71} & \textbf{151} & \textbf{144} & \textbf{71} & 171 & 160 \\
\texttt{rc\_adder\_6} & 14 & 93 & \textbf{71} & \textbf{102} & \textbf{102} & 81 & 119 & 119 \\
\texttt{gf2\textasciicircum{}E4\_mult} & 12 & 99 & \textbf{99} & \textbf{228} & \textbf{202} & \textbf{99} & 241 & 228 \\
\texttt{tof\_10} & 19 & 102 & \textbf{70} & \textbf{101} & \textbf{101} & 90 & 107 & 105 \\
\texttt{mod\_red\_21} & 11 & 105 & \textbf{81} & \textbf{153} & \textbf{149} & 99 & 208 & 198 \\
\texttt{gf2\textasciicircum{}E5\_mult} & 15 & 154 & \textbf{154} & \textbf{400} & \textbf{355} & \textbf{154} & 441 & 431 \\
\texttt{csum\_mux\_9\_corrected} & 30 & 168 & \textbf{140} & \textbf{342} & \textbf{318} & 164 & 392 & 375 \\
\texttt{QFTAdd8} & 16 & 184 & \textbf{176} & \textbf{363} & \textbf{342} & 183 & 477 & 435 \\
\texttt{qcla\_com\_7} & 24 & 186 & \textbf{143} & \textbf{321} & \textbf{301} & 156 & 334 & 328 \\
\texttt{barenco\_tof\_10} & 19 & 192 & \textbf{130} & \textbf{178} & \textbf{178} & 182 & 216 & 187 \\
\texttt{gf2\textasciicircum{}E6\_mult} & 18 & 221 & \textbf{221} & \textbf{580} & \textbf{531} & \textbf{221} & 665 & 628 \\
\texttt{QFT16} & 16 & 228 & \textbf{220} & \textbf{356} & \textbf{341} & 228 & 449 & 392 \\
\texttt{qcla\_adder\_10} & 36 & 233 & \textbf{209} & \textbf{456} & \textbf{429} & 211 & 470 & 445 \\
\texttt{gf2\textasciicircum{}E7\_mult} & 21 & 300 & \textbf{300} & \textbf{863} & \textbf{772} & \textbf{300} & 936 & 875 \\
\texttt{qcla\_mod\_7} & 26 & 382 & \textbf{350} & \textbf{907} & \textbf{857} & 352 & 1054 & 998 \\
\texttt{gf2\textasciicircum{}E8\_mult} & 24 & 405 & \textbf{405} & \textbf{1236} & \textbf{1106} & \textbf{405} & 1387 & 1316 \\
\texttt{adder\_8} & 24 & 409 & \textbf{298} & \textbf{639} & \textbf{588} & 377 & 653 & 606 \\
\texttt{gf2\textasciicircum{}E9\_mult} & 27 & 494 & \textbf{494} & \textbf{1499} & \textbf{1354} & \textbf{494} & 1956 & 1878 \\
\texttt{gf2\textasciicircum{}E10\_mult} & 30 & 609 & \textbf{609} & \textbf{1985} & \textbf{1787} & \textbf{609} & 2846 & 2720 \\
\texttt{QFT32} & 32 & 612 & \textbf{596} & \textbf{1484} & \textbf{1465} & 612 & 1827 & 1721 \\
\texttt{QFTAdd16} & 32 & 716 & \textbf{700} & \textbf{1602} & \textbf{1507} & 716 & 2102 & 1897 \\
\texttt{QFT64} & 64 & 1380 & \textbf{1346} & \textbf{3718} & \textbf{3568} & 1380 & 7570 & 7241 \\
\texttt{gf2\textasciicircum{}E16\_mult} & 48 & 1581 & \textbf{1581} & \textbf{6234} & \textbf{5787} & \textbf{1581} & 6867 & 6597 \\
\texttt{mod\_adder\_1024} & 28 & 1720 & \textbf{1236} & \textbf{2110} & \textbf{2040} & 1611 & 2446 & 2229 \\
\texttt{QFTAdd32} & 64 & 1900 & \textbf{1868} & \textbf{7477} & \textbf{7002} & 1899 & 8680 & 8136 \\
\hline
Total & -- & 12989 & \textbf{11894} & \textbf{33943} & \textbf{31775} & 12668 & 43153 & 40748 \\
\hline
\end{tabular}
\caption{IBM Eagle three-qubit partition CNOT counts for Squander and BQSKit at matched partition resynthesis infidelity tolerance. Bold values are best within each paired stage. Refreshed completed-result cohort: Squander has 59 completed circuits at each partition size (35 IBM Eagle and 24 QASMBenchmarks). Paired comparisons use only circuits completed by both workflows; Squander-only four-qubit results are listed separately. QFT128 has no completed Squander result at this cutoff and is not counted as a failure.}
\label{tab:osr-bqskit-three-qubit-ibm-eagle-counts}
\end{table*}

We tested circuits from QASMBench~\cite{Li2023QASMBench} and the IBM-EAGLE benchmark set used in the GUOQ evaluation~\cite{Xu2025GUOQ}. Before synthesis, we converted each circuit to the $(u3,\mathrm{CX})$ basis. The \texttt{tfim} circuit present in the tables is from QFAST/QSearch benchmark \texttt{tfim-4-60}~\cite{Younis2021QFAST}, not a member of the original QASMBench suite. The tables report CNOT counts and stage runtimes for maximum partition sizes of three and four qubits. Routing uses blocks of at most three qubits in both settings. We include only completed and verified results. The availability audit records missing results, failures, and timeouts.
{ These runtimes compare complete workflows with different search scopes and should be interpreted separately by stage.} Squander enumerates all the admissible partition candidates within the configured block-size limit and uses an ILP to select a globally optimal cover over those candidates and their computed costs, whereas the BQSKit workflow operates on partitions produced by its heuristic approach. { Routing includes constructing the block/permutation synthesis library and searching over mappings and partitionings. The refreshed Routing and Final times are cumulative experiment costs, including the archived runs and additional search and resynthesis work.}

We validate the end-to-end result by numerically checking accepted local rewrites (i.e. partition resynthesis) and checking the ordered sequence of partition covers, basis conversions, and routing mappings. The exported gates and parameters must match this validated sequence. For large circuits, this compositional check avoids simulating the full circuit, which would be otherwise highly demanding.

\begin{table*}[t]
\centering
\begin{tabular}{l|rrr|rrr}
\hline
Circuit & \multicolumn{3}{c|}{Squander time (min)} & \multicolumn{3}{c}{BQSKit time (min)} \\
 & A2A & Routing & Final & A2A & Routing & Final \\
\hline
\texttt{tof\_3} & \textbf{0.05} & 0.84 & \textbf{0.04} & 0.84 & \textbf{0.52} & 0.47 \\
\texttt{barenco\_tof\_3} & \textbf{0.05} & 0.82 & \textbf{0.06} & 1.08 & \textbf{0.57} & 0.43 \\
\texttt{mod5\_4} & \textbf{0.11} & 0.92 & \textbf{0.07} & 1.11 & \textbf{0.51} & 0.79 \\
\texttt{tof\_4} & \textbf{0.04} & 0.96 & \textbf{0.18} & 1.08 & \textbf{0.59} & 0.55 \\
\texttt{tof\_5} & \textbf{0.05} & 1.25 & \textbf{0.22} & 1.07 & \textbf{0.71} & 0.61 \\
\texttt{barenco\_tof\_4} & \textbf{0.11} & 1.38 & \textbf{0.16} & 1.07 & \textbf{0.66} & 0.69 \\
\texttt{mod\_mult\_55} & \textbf{0.14} & 2.03 & \textbf{0.55} & 0.80 & \textbf{0.76} & 1.02 \\
\texttt{QFT8} & \textbf{0.31} & 2.59 & \textbf{0.56} & 0.59 & \textbf{0.89} & 1.28 \\
\texttt{vbe\_adder\_3} & \textbf{0.35} & 2.20 & \textbf{0.24} & 0.74 & \textbf{0.37} & 0.43 \\
\texttt{barenco\_tof\_5} & \textbf{0.24} & 1.94 & \textbf{0.22} & 1.14 & \textbf{0.99} & 0.90 \\
\texttt{csla\_mux\_3\_original} & \textbf{0.25} & 3.99 & 1.89 & 1.10 & \textbf{1.67} & \textbf{1.07} \\
\texttt{rc\_adder\_6} & \textbf{0.10} & 4.27 & \textbf{0.24} & 0.76 & \textbf{0.97} & 0.71 \\
\texttt{gf2\textasciicircum{}E4\_mult} & \textbf{0.04} & 4.11 & 2.55 & 0.87 & \textbf{1.35} & \textbf{1.25} \\
\texttt{tof\_10} & \textbf{0.08} & 3.14 & \textbf{0.23} & 1.12 & \textbf{1.07} & 0.59 \\
\texttt{mod\_red\_21} & \textbf{0.11} & 4.02 & 1.47 & 1.49 & \textbf{1.23} & \textbf{1.43} \\
\texttt{gf2\textasciicircum{}E5\_mult} & \textbf{0.04} & 8.07 & 4.70 & 0.96 & \textbf{1.94} & \textbf{1.32} \\
\texttt{csum\_mux\_9\_corrected} & \textbf{0.15} & 12.29 & 2.99 & 1.35 & \textbf{2.00} & \textbf{1.90} \\
\texttt{QFTAdd8} & \textbf{0.73} & 10.62 & 3.32 & 1.19 & \textbf{2.09} & \textbf{2.25} \\
\texttt{qcla\_com\_7} & \textbf{0.25} & 9.89 & 2.17 & 1.53 & \textbf{1.65} & \textbf{1.09} \\
\texttt{barenco\_tof\_10} & \textbf{0.17} & 5.54 & \textbf{0.18} & 1.79 & \textbf{2.12} & 0.91 \\
\texttt{gf2\textasciicircum{}E6\_mult} & \textbf{0.05} & 10.35 & 8.19 & 1.11 & \textbf{2.75} & \textbf{2.10} \\
\texttt{QFT16} & \textbf{0.58} & 13.15 & 4.43 & 0.81 & \textbf{2.18} & \textbf{2.39} \\
\texttt{qcla\_adder\_10} & \textbf{0.26} & 18.92 & 4.94 & 0.96 & \textbf{2.33} & \textbf{1.57} \\
\texttt{gf2\textasciicircum{}E7\_mult} & \textbf{0.06} & 14.84 & 16.90 & 1.31 & \textbf{3.66} & \textbf{2.70} \\
\texttt{qcla\_mod\_7} & \textbf{0.44} & 23.98 & 12.58 & 1.27 & \textbf{3.74} & \textbf{3.12} \\
\texttt{gf2\textasciicircum{}E8\_mult} & \textbf{0.06} & 21.21 & 19.19 & 1.63 & \textbf{4.80} & \textbf{3.16} \\
\texttt{adder\_8} & \textbf{0.70} & 19.04 & 6.46 & 1.70 & \textbf{3.75} & \textbf{2.18} \\
\texttt{gf2\textasciicircum{}E9\_mult} & \textbf{0.07} & 30.77 & 25.20 & 1.98 & \textbf{5.92} & \textbf{5.94} \\
\texttt{gf2\textasciicircum{}E10\_mult} & \textbf{0.08} & 45.74 & 94.74 & 2.35 & \textbf{7.31} & \textbf{8.83} \\
\texttt{QFT32} & \textbf{1.02} & 39.78 & 29.81 & 1.52 & \textbf{5.77} & \textbf{7.91} \\
\texttt{QFTAdd16} & 2.11 & 46.14 & 886.98 & \textbf{1.67} & \textbf{6.92} & \textbf{9.52} \\
\texttt{QFT64} & 3.10 & 130.78 & 1204.49 & \textbf{2.66} & \textbf{12.76} & \textbf{28.04} \\
\texttt{gf2\textasciicircum{}E16\_mult} & \textbf{0.17} & 108.57 & 1611.62 & 5.99 & \textbf{19.38} & \textbf{19.04} \\
\texttt{mod\_adder\_1024} & \textbf{8.48} & 895.67 & 57.95 & 8.51 & \textbf{17.15} & \textbf{6.55} \\
\texttt{QFTAdd32} & \textbf{4.81} & 196.14 & 1932.09 & 5.79 & \textbf{17.53} & \textbf{48.43} \\
\hline
Total & \textbf{25.38} & 1695.94 & 5937.59 & 60.95 & \textbf{138.61} & \textbf{171.19} \\
\hline
\end{tabular}
\caption{IBM Eagle three-qubit partition stage runtimes for Squander and BQSKit at matched partition resynthesis infidelity tolerance, in minutes. Bold values are best within each paired stage.}
\label{tab:osr-bqskit-three-qubit-ibm-eagle-times}
\end{table*}

Across the 35 matched IBM Eagle circuits optimized using three-qubit partitions (Tables~\ref{tab:osr-bqskit-three-qubit-ibm-eagle-counts} and~\ref{tab:osr-bqskit-three-qubit-ibm-eagle-times}), Squander's CNOT advantage is evident at all three stages. Paired tests with BQSKit confirm lower counts at all three stages. Squander wins all 35 Routed and all 35 Final comparisons. The CNOT advantage therefore survives routing on this cohort. Runtime results show a similar stage-dependent trade-off: Squander is usually faster in A2A, but cumulative routing times are higher in every pair and a few long Squander runs dominate the total. The largest runtime gap appears in Final, where Squander's total is much higher with a higher per-circuit median time as well. Overall, Squander generally improves final CNOT counts, but runtime is less predictable and can be dominated by a small number of expensive cases. Summed over these 35 matched circuits, Squander uses 774 fewer CNOTs in A2A, 9,210 fewer after routing, and 8,973 fewer after Final optimization than BQSKit.

\begin{table*}[t]
\centering
\begin{tabular}{lrr|rrr|rrr}
\hline
Circuit & Qubits & Initial CNOTs & \multicolumn{3}{c|}{Squander CNOTs} & \multicolumn{3}{c}{BQSKit CNOTs} \\
 & & & A2A & Routed & Final & A2A & Routed & Final \\
\hline
\texttt{adder\_n4} & 4 & 10 & \textbf{10} & \textbf{12} & \textbf{10} & \textbf{10} & 16 & 16 \\
\texttt{bv\_n14} & 14 & 13 & \textbf{13} & \textbf{24} & \textbf{24} & \textbf{13} & 27 & \textbf{24} \\
\texttt{cat\_state\_n22} & 22 & 21 & -- & -- & \textbf{21} & -- & -- & \textbf{21} \\
\texttt{ghz\_state\_n23} & 23 & 22 & -- & -- & \textbf{22} & -- & -- & \textbf{22} \\
\texttt{multiply\_n13} & 13 & 40 & \textbf{40} & \textbf{79} & \textbf{77} & \textbf{40} & 80 & 78 \\
\texttt{tfim} & 4 & 60 & -- & -- & \textbf{29} & -- & -- & 56 \\
\texttt{bv\_n140} & 140 & 72 & \textbf{72} & \textbf{142} & \textbf{142} & \textbf{72} & 348 & 334 \\
\texttt{swap\_test\_n25} & 25 & 84 & \textbf{84} & \textbf{106} & \textbf{106} & \textbf{84} & \textbf{106} & \textbf{106} \\
\texttt{ising\_n10} & 10 & 90 & -- & -- & \textbf{50} & -- & -- & 76 \\
\texttt{dnn\_n8} & 8 & 192 & \textbf{64} & \textbf{108} & \textbf{98} & \textbf{64} & 134 & 126 \\
\texttt{grover\_19} & 19 & 192 & \textbf{124} & \textbf{182} & \textbf{179} & 165 & 204 & 198 \\
\texttt{ising\_n98} & 98 & 194 & -- & -- & \textbf{194} & -- & -- & \textbf{194} \\
\texttt{hhl\_n7} & 7 & 196 & \textbf{86} & \textbf{158} & \textbf{149} & 89 & 197 & 192 \\
\texttt{grover\_23} & 23 & 240 & \textbf{154} & \textbf{229} & \textbf{225} & 218 & 262 & 244 \\
\texttt{multiplier\_n15} & 15 & 246 & \textbf{184} & \textbf{387} & \textbf{348} & 210 & 450 & 418 \\
\texttt{heisenberg-16-20} & 16 & 360 & -- & -- & \textbf{193} & -- & -- & 204 \\
\texttt{grover\_33} & 33 & 360 & \textbf{229} & \textbf{345} & \textbf{333} & 322 & 389 & 371 \\
\texttt{dnn\_n16} & 16 & 384 & \textbf{128} & \textbf{233} & \textbf{224} & \textbf{128} & 314 & 276 \\
\texttt{qft\_20} & 20 & 410 & \textbf{387} & \textbf{854} & \textbf{809} & 389 & 992 & 908 \\
\texttt{adder\_n64} & 64 & 455 & \textbf{287} & \textbf{345} & \textbf{345} & 343 & 2400 & 2366 \\
\texttt{grover\_45} & 45 & 504 & \textbf{319} & \textbf{462} & \textbf{460} & 429 & 530 & 510 \\
\texttt{grover\_61} & 61 & 696 & \textbf{439} & \textbf{635} & \textbf{633} & 613 & 744 & 705 \\
\texttt{grover\_81} & 81 & 936 & \textbf{589} & \textbf{859} & \textbf{857} & 824 & 999 & 955 \\
\texttt{quantum\_volume\_n40\_d20\_i0} & 40 & 1200 & \textbf{1170} & \textbf{9264} & \textbf{8076} & \textbf{1170} & 11797 & 11070 \\
\hline
Total & -- & 6977 & \textbf{4379} & \textbf{14424} & \textbf{13604} & 5183 & 19989 & 19470 \\
\hline
\end{tabular}
\caption{QASMBenchmarks three-qubit partition CNOT counts for Squander and BQSKit at matched partition resynthesis infidelity tolerance. Bold values are best within each paired stage. The tfim row is the QFAST/QSearch tfim-4-60 benchmark; it is not an original QASMBenchmarks circuit.}
\label{tab:osr-bqskit-three-qubit-qasm-benchmarks-counts}
\end{table*}

\begin{table*}[t]
\centering
\begin{tabular}{l|rrr|rrr}
\hline
Circuit & \multicolumn{3}{c|}{Squander time (min)} & \multicolumn{3}{c}{BQSKit time (min)} \\
 & A2A & Routing & Final & A2A & Routing & Final \\
\hline
\texttt{adder\_n4} & \textbf{0.02} & 0.58 & \textbf{0.04} & 0.14 & \textbf{0.14} & 0.39 \\
\texttt{bv\_n14} & \textbf{0.00} & 0.95 & \textbf{0.00} & 0.12 & \textbf{0.22} & 0.21 \\
\texttt{cat\_state\_n22} & -- & -- & \textbf{0.00} & -- & -- & 0.11 \\
\texttt{ghz\_state\_n23} & -- & -- & \textbf{0.00} & -- & -- & 0.12 \\
\texttt{multiply\_n13} & \textbf{0.04} & 2.12 & \textbf{0.89} & 0.81 & \textbf{0.70} & 1.10 \\
\texttt{tfim} & -- & -- & 3.84 & -- & -- & \textbf{0.57} \\
\texttt{bv\_n140} & \textbf{0.01} & 17.78 & 1.90 & 0.18 & \textbf{0.72} & \textbf{0.71} \\
\texttt{swap\_test\_n25} & \textbf{0.10} & 4.65 & 2.13 & 1.21 & \textbf{4.05} & \textbf{0.95} \\
\texttt{ising\_n10} & -- & -- & 1.03 & -- & -- & \textbf{0.39} \\
\texttt{dnn\_n8} & 5.05 & 4.86 & 1.52 & \textbf{0.87} & \textbf{3.41} & \textbf{1.16} \\
\texttt{grover\_19} & \textbf{0.36} & 7.80 & 1.35 & 2.01 & \textbf{2.25} & \textbf{1.11} \\
\texttt{ising\_n98} & -- & -- & \textbf{0.17} & -- & -- & 0.29 \\
\texttt{hhl\_n7} & 1.53 & 4.18 & 1.57 & \textbf{1.19} & \textbf{0.94} & \textbf{0.75} \\
\texttt{grover\_23} & \textbf{0.46} & 9.92 & 1.52 & 1.90 & \textbf{2.39} & \textbf{1.23} \\
\texttt{multiplier\_n15} & \textbf{0.34} & 10.35 & 3.26 & 1.78 & \textbf{2.13} & \textbf{2.02} \\
\texttt{heisenberg-16-20} & -- & -- & 1.49 & -- & -- & \textbf{1.17} \\
\texttt{grover\_33} & \textbf{0.80} & 16.17 & 2.27 & 2.79 & \textbf{3.94} & \textbf{0.81} \\
\texttt{dnn\_n16} & 6.92 & 8.81 & \textbf{0.61} & \textbf{0.93} & \textbf{5.06} & 6.61 \\
\texttt{qft\_20} & 1.58 & 26.17 & 11.31 & \textbf{1.15} & \textbf{3.37} & \textbf{3.75} \\
\texttt{adder\_n64} & 1.75 & 27.28 & 4.30 & \textbf{1.58} & \textbf{3.58} & \textbf{3.48} \\
\texttt{grover\_45} & \textbf{1.41} & 24.25 & 2.92 & 4.02 & \textbf{4.39} & \textbf{1.44} \\
\texttt{grover\_61} & \textbf{2.56} & 38.92 & 4.39 & 5.45 & \textbf{6.00} & \textbf{1.86} \\
\texttt{grover\_81} & \textbf{4.57} & 66.90 & 6.97 & 7.05 & \textbf{7.81} & \textbf{2.11} \\
\texttt{quantum\_volume\_n40\_d20\_i0} & 70.15 & 226.52 & 602.62 & \textbf{2.38} & \textbf{92.26} & \textbf{50.72} \\
\hline
Total & 97.64 & 498.20 & 656.10 & \textbf{35.57} & \textbf{143.34} & \textbf{83.05} \\
\hline
\end{tabular}
\caption{QASMBenchmarks three-qubit partition stage runtimes for Squander and BQSKit at matched partition resynthesis infidelity tolerance, in minutes. Bold values are best within each paired stage. The tfim row is the QFAST/QSearch tfim-4-60 benchmark; it is not an original QASMBenchmarks circuit.}
\label{tab:osr-bqskit-three-qubit-qasm-benchmarks-times}
\end{table*}

For the Squander-BQSKit matched QASMBenchmarks circuits optimized using three-qubit partitions (Tables~\ref{tab:osr-bqskit-three-qubit-qasm-benchmarks-counts} and~\ref{tab:osr-bqskit-three-qubit-qasm-benchmarks-times}), Squander reduces CNOT counts across A2A, Routed, and Final, with paired differences significant at every stage. As on IBM Eagle, the CNOT advantage survives routing on this benchmark set. These results describe the completed cohort: some circuits tie, and none favors BQSKit in CNOT count. Runtime tells a more nuanced story: Squander's median A2A time is slightly lower, but its Routing and Final medians are higher, and BQSKit's cumulative time is lower at every stage. Thus, three-qubit partitions provide a statistically supported CNOT advantage on QASMBenchmarks, while the median runtime benefit is limited to A2A. The table totals show 804 fewer Squander CNOTs in A2A and 5,565 fewer after routing across the 18 paired circuits with counts at those stages, plus 5,866 fewer at Final across 24 paired circuits.

These results motivate our next experiment with four-qubit partitions. Larger blocks let resynthesis consider more gates together and may therefore capture additional opportunities to reduce CNOT counts, at the cost of increased synthesis time. The following results evaluate this expected compression--runtime trade-off.

\begin{table*}[t]
\centering
\begin{tabular}{lrr|rrr|rrr}
\hline
Circuit & Qubits & Initial CNOTs & \multicolumn{3}{c|}{Squander CNOTs} & \multicolumn{3}{c}{BQSKit CNOTs} \\
 & & & A2A & Routed & Final & A2A & Routed & Final \\
\hline
\texttt{tof\_3} & 5 & 18 & \textbf{14} & \textbf{17} & \textbf{17} & 18 & 19 & 19 \\
\texttt{barenco\_tof\_3} & 5 & 24 & \textbf{18} & \textbf{23} & \textbf{23} & 24 & 27 & 27 \\
\texttt{mod5\_4} & 5 & 28 & 16 & \textbf{20} & \textbf{18} & \textbf{12} & 23 & 19 \\
\texttt{tof\_4} & 7 & 30 & \textbf{22} & \textbf{29} & \textbf{29} & 30 & 36 & 34 \\
\texttt{tof\_5} & 9 & 42 & \textbf{30} & \textbf{41} & \textbf{41} & 41 & 53 & 48 \\
\texttt{barenco\_tof\_4} & 7 & 48 & \textbf{34} & \textbf{46} & \textbf{46} & 44 & 52 & 50 \\
\texttt{mod\_mult\_55} & 9 & 48 & \textbf{34} & \textbf{81} & 74 & 40 & 95 & \textbf{68} \\
\texttt{vbe\_adder\_3} & 10 & 70 & \textbf{38} & \textbf{60} & \textbf{52} & \textbf{38} & 75 & 57 \\
\texttt{barenco\_tof\_5} & 9 & 72 & \textbf{50} & \textbf{68} & \textbf{68} & 65 & 97 & 75 \\
\texttt{csla\_mux\_3\_original} & 15 & 80 & \textbf{68} & \textbf{155} & \textbf{139} & 70 & 188 & 160 \\
\texttt{rc\_adder\_6} & 14 & 93 & \textbf{71} & \textbf{102} & 97 & 79 & 117 & \textbf{96} \\
\texttt{gf2\textasciicircum{}E4\_mult} & 12 & 99 & \textbf{91} & \textbf{248} & \textbf{215} & 99 & 296 & 260 \\
\texttt{tof\_10} & 19 & 102 & \textbf{70} & \textbf{101} & \textbf{101} & 96 & 123 & 117 \\
\texttt{gf2\textasciicircum{}E5\_mult} & 15 & 154 & \textbf{149} & \textbf{426} & \textbf{361} & 154 & 583 & 541 \\
\texttt{csum\_mux\_9\_corrected} & 30 & 168 & \textbf{126} & \textbf{345} & \textbf{289} & 132 & 446 & 417 \\
\texttt{qcla\_com\_7} & 24 & 186 & \textbf{140} & \textbf{317} & \textbf{280} & 155 & 415 & 347 \\
\texttt{barenco\_tof\_10} & 19 & 192 & \textbf{130} & \textbf{178} & \textbf{178} & 186 & 263 & 244 \\
\texttt{gf2\textasciicircum{}E6\_mult} & 18 & 221 & \textbf{203} & \textbf{651} & \textbf{577} & 221 & 733 & 656 \\
\texttt{qcla\_adder\_10} & 36 & 233 & \textbf{200} & \textbf{465} & \textbf{403} & 205 & 470 & 432 \\
\texttt{adder\_8} & 24 & 409 & \textbf{243} & \textbf{564} & \textbf{491} & 251 & 650 & 578 \\
\hline
Total & -- & 2317 & \textbf{1747} & \textbf{3937} & \textbf{3499} & 1960 & 4761 & 4245 \\
\hline
\end{tabular}
\caption{IBM Eagle four-qubit partition CNOT counts for Squander and BQSKit at matched partition resynthesis infidelity tolerance. Bold values are best within each paired stage.}
\label{tab:osr-bqskit-four-qubit-ibm-eagle-counts}
\end{table*}

\begin{table*}[t]
\centering
\begin{tabular}{l|rrr|rrr}
\hline
Circuit & \multicolumn{3}{c|}{Squander time (min)} & \multicolumn{3}{c}{BQSKit time (min)} \\
 & A2A & Routing & Final & A2A & Routing & Final \\
\hline
\texttt{tof\_3} & \textbf{0.24} & 0.96 & \textbf{0.38} & 17.67 & \textbf{0.38} & 32.95 \\
\texttt{barenco\_tof\_3} & \textbf{0.26} & 1.18 & \textbf{0.41} & 28.08 & \textbf{0.50} & 5.94 \\
\texttt{mod5\_4} & \textbf{0.79} & 1.30 & \textbf{0.27} & 290.28 & \textbf{0.22} & 25.15 \\
\texttt{tof\_4} & \textbf{0.39} & 1.28 & \textbf{0.82} & 31.02 & \textbf{1.06} & 31.25 \\
\texttt{tof\_5} & \textbf{0.51} & 1.56 & \textbf{1.17} & 110.37 & \textbf{0.71} & 14.89 \\
\texttt{barenco\_tof\_4} & \textbf{0.67} & 1.92 & \textbf{0.65} & 22.92 & \textbf{0.68} & 8.37 \\
\texttt{mod\_mult\_55} & \textbf{1.98} & 2.87 & \textbf{3.49} & 866.15 & \textbf{0.52} & 120.32 \\
\texttt{vbe\_adder\_3} & \textbf{1.01} & 2.93 & \textbf{2.62} & 387.52 & \textbf{0.43} & 80.95 \\
\texttt{barenco\_tof\_5} & \textbf{0.88} & 2.56 & \textbf{0.93} & 71.18 & \textbf{0.80} & 79.25 \\
\texttt{csla\_mux\_3\_original} & \textbf{4.23} & 6.06 & \textbf{14.57} & 69.07 & \textbf{1.55} & 93.58 \\
\texttt{rc\_adder\_6} & \textbf{0.72} & 5.12 & \textbf{1.79} & 63.24 & \textbf{0.83} & 36.32 \\
\texttt{gf2\textasciicircum{}E4\_mult} & \textbf{4.26} & 8.04 & \textbf{8.71} & 35.86 & \textbf{1.23} & 190.22 \\
\texttt{tof\_10} & \textbf{0.47} & 3.55 & \textbf{1.25} & 186.80 & \textbf{1.20} & 35.41 \\
\texttt{gf2\textasciicircum{}E5\_mult} & \textbf{3.82} & 10.31 & \textbf{26.04} & 160.07 & \textbf{1.72} & 484.14 \\
\texttt{csum\_mux\_9\_corrected} & \textbf{1.49} & 11.83 & \textbf{17.61} & 423.66 & \textbf{1.37} & 373.43 \\
\texttt{qcla\_com\_7} & \textbf{2.98} & 11.71 & \textbf{9.77} & 356.93 & \textbf{1.52} & 208.98 \\
\texttt{barenco\_tof\_10} & \textbf{0.81} & 6.33 & \textbf{0.99} & 39.69 & \textbf{1.98} & 49.16 \\
\texttt{gf2\textasciicircum{}E6\_mult} & \textbf{4.97} & 16.51 & \textbf{36.79} & 40.22 & \textbf{2.75} & 470.18 \\
\texttt{qcla\_adder\_10} & \textbf{4.31} & 23.37 & \textbf{28.68} & 220.84 & \textbf{2.20} & 380.15 \\
\texttt{adder\_8} & \textbf{9.66} & 24.30 & \textbf{32.79} & 353.28 & \textbf{2.48} & 181.86 \\
\hline
Total & \textbf{44.44} & 143.69 & \textbf{189.71} & 3774.84 & \textbf{24.12} & 2902.52 \\
\hline
\end{tabular}
\caption{IBM Eagle four-qubit partition stage runtimes for Squander and BQSKit at matched partition resynthesis infidelity tolerance, in minutes. Bold values are best within each paired stage.}
\label{tab:osr-bqskit-four-qubit-ibm-eagle-times}
\end{table*}

On the 20 IBM Eagle circuits with paired four-qubit results (Tables~\ref{tab:osr-bqskit-four-qubit-ibm-eagle-counts} and~\ref{tab:osr-bqskit-four-qubit-ibm-eagle-times}), Squander achieves significantly lower CNOT counts than BQSKit at every stage. It reduces the count in most circuits, and the advantage remains evident after routing and final optimization; it wins all 20 Routed comparisons and 18 of the 20 Final comparisons. Comparing Squander on the same circuits at the two partition sizes, four-qubit partitions reduce aggregate A2A and Final CNOT counts. This additional compression comes at a substantial runtime cost relative to three-qubit Squander. Against BQSKit, however, four-qubit Squander is much faster overall in the A2A and Final stages, while recorded routing time is higher for Squander in all 20 paired cases. These results show that larger partitions strengthen the CNOT outcome and can reduce the total optimization time relative to BQSKit in two stages. Across the 20 matched circuits, Squander uses 213 fewer CNOTs in A2A, 824 fewer after routing, and 746 fewer after Final optimization than BQSKit.

\begin{table*}[t]
\centering
\begin{tabular}{lrr|rr|rr|rr}
\hline
Circuit & Qubits & Initial CNOTs & \multicolumn{2}{c|}{A2A} & \multicolumn{2}{c|}{Routed} & \multicolumn{2}{c}{Final} \\
 & & & CNOTs & Time (min) & CNOTs & Time (min) & CNOTs & Time (min) \\
\hline
\texttt{gf2\textasciicircum{}E9\_mult} & 27 & 494 & 468 & 6.19 & 1653 & 38.29 & 1373 & 936.74 \\
\texttt{QFT8} & 8 & 56 & 51 & 15.52 & 73 & 4.49 & 73 & 5.71 \\
\texttt{mod\_red\_21} & 11 & 105 & 78 & 2.08 & 156 & 5.96 & 148 & 8.09 \\
\texttt{QFTAdd8} & 16 & 184 & 173 & 39.50 & 377 & 13.92 & 352 & 33.97 \\
\texttt{QFT16} & 16 & 228 & 220 & 26.85 & 362 & 14.91 & 343 & 40.65 \\
\texttt{gf2\textasciicircum{}E7\_mult} & 21 & 300 & 279 & 7.25 & 926 & 22.45 & 804 & 49.69 \\
\texttt{qcla\_mod\_7} & 26 & 382 & 345 & 7.54 & 917 & 29.73 & 809 & 63.10 \\
\texttt{gf2\textasciicircum{}E8\_mult} & 24 & 405 & 381 & 3.74 & 1292 & 27.27 & 1115 & 72.78 \\
\texttt{gf2\textasciicircum{}E10\_mult} & 30 & 609 & 577 & 7.80 & 2325 & 108.57 & 1974 & 1005.80 \\
\texttt{QFT32} & 32 & 612 & 596 & 27.39 & 1467 & 48.14 & 1413 & 151.72 \\
\texttt{QFTAdd16} & 32 & 716 & 673 & 129.87 & 1957 & 231.79 & 1633 & 1050.47 \\
\texttt{QFT64} & 64 & 1380 & 1348 & 29.82 & 3645 & 128.86 & 3431 & 1613.74 \\
\texttt{gf2\textasciicircum{}E16\_mult} & 48 & 1581 & 1492 & 14.41 & 6820 & 123.89 & 5996 & 1665.33 \\
\texttt{mod\_adder\_1024} & 28 & 1720 & 1124 & 22.93 & 2476 & 73.14 & 2131 & 1041.08 \\
\texttt{QFTAdd32} & 64 & 1900 & 1787 & 143.74 & 8807 & 178.55 & 7535 & 2062.94 \\
\hline
Total & -- & 10672 & 9592 & 484.63 & 33253 & 1049.95 & 29130 & 9801.83 \\
\hline
\end{tabular}
\caption{Completed IBM Eagle four-qubit partition Squander results for circuits without a completed BQSKit four-qubit result. Counts and stage runtimes are combined because no paired comparison is available.}
\label{tab:osr-four-qubit-unpaired-ibm-eagle}
\end{table*}
Considering all 35 IBM Eagle circuits with Squander results at both partition sizes gives a more qualified picture than the 20-circuit paired comparison above. In this broader Squander-only comparison, four-qubit partitions reduce the aggregate A2A count (11,339 versus 11,894 for three-qubit partitions), but the routed count is higher (37,190 versus 33,943), and the final total is also higher (32,629 versus 31,775). Thus, the A2A savings do not carry through to the Routed or Final totals. 

The shift between A2A, Routed, and Final counts shows that routing is not a secondary implementation detail: hardware connectivity and routing choices can substantially reshape, or erase, gains achieved by block resynthesis. In the full cohort above, four-qubit max partitions lower the A2A count but yield higher routed and final totals than three-qubit partitions. Thus, resynthesis cost alone does not determine the quality of an end-to-end circuit. The main goal of this work, however, is to advance large-circuit optimization by enumerating the full set of admissible partitions under a prescribed block-size limit and using an ILP to select the globally optimal cover over those candidates and their computed costs. Pairing this broader search with OSR-based resynthesis of smaller blocks makes the approach practically tractable. These results highlight routing as an important direction for further improvement while underscoring the central contribution of scalable global partition selection.

\begin{table*}[t]
\centering
\begin{tabular}{lrr|rrr|rrr}
\hline
Circuit & Qubits & Initial CNOTs & \multicolumn{3}{c|}{Squander CNOTs} & \multicolumn{3}{c}{BQSKit CNOTs} \\
 & & & A2A & Routed & Final & A2A & Routed & Final \\
\hline
\texttt{adder\_n4} & 4 & 10 & \textbf{7} & \textbf{9} & \textbf{9} & \textbf{7} & 14 & 10 \\
\texttt{bv\_n14} & 14 & 13 & \textbf{13} & \textbf{24} & \textbf{24} & \textbf{13} & 25 & \textbf{24} \\
\texttt{cat\_state\_n22} & 22 & 21 & -- & -- & \textbf{21} & -- & -- & \textbf{21} \\
\texttt{ghz\_state\_n23} & 23 & 22 & -- & -- & \textbf{22} & -- & -- & \textbf{22} \\
\texttt{multiply\_n13} & 13 & 40 & \textbf{40} & \textbf{79} & \textbf{72} & \textbf{40} & 146 & 121 \\
\texttt{tfim} & 4 & 60 & -- & -- & \textbf{12} & -- & -- & \textbf{12} \\
\texttt{bv\_n140} & 140 & 72 & \textbf{72} & \textbf{142} & \textbf{142} & \textbf{72} & 345 & 334 \\
\texttt{swap\_test\_n25} & 25 & 84 & \textbf{84} & \textbf{106} & \textbf{106} & \textbf{84} & \textbf{106} & \textbf{106} \\
\texttt{ising\_n10} & 10 & 90 & -- & -- & \textbf{44} & -- & -- & 58 \\
\texttt{grover\_19} & 19 & 192 & \textbf{124} & \textbf{182} & \textbf{179} & 177 & 224 & 201 \\
\texttt{ising\_n98} & 98 & 194 & -- & -- & \textbf{194} & -- & -- & \textbf{194} \\
\texttt{hhl\_n7} & 7 & 196 & \textbf{84} & \textbf{158} & \textbf{149} & 87 & 201 & 179 \\
\texttt{grover\_23} & 23 & 240 & \textbf{154} & \textbf{229} & \textbf{225} & 218 & 286 & 271 \\
\texttt{multiplier\_n15} & 15 & 246 & \textbf{138} & \textbf{332} & \textbf{267} & 142 & 503 & 396 \\
\texttt{heisenberg-16-20} & 16 & 360 & -- & -- & \textbf{138} & -- & -- & 157 \\
\texttt{grover\_33} & 33 & 360 & \textbf{229} & \textbf{345} & \textbf{333} & 336 & 419 & 390 \\
\texttt{grover\_45} & 45 & 504 & \textbf{319} & \textbf{462} & \textbf{458} & 476 & 789 & 745 \\
\texttt{grover\_61} & 61 & 696 & \textbf{439} & \textbf{635} & \textbf{631} & 647 & 815 & 768 \\
\hline
Total & -- & 3400 & \textbf{1703} & \textbf{2703} & \textbf{3026} & 2299 & 3873 & 4009 \\
\hline
\end{tabular}
\caption{QASMBenchmarks four-qubit partition CNOT counts for Squander and BQSKit at matched partition resynthesis infidelity tolerance. Bold values are best within each paired stage. The tfim row is the QFAST/QSearch tfim-4-60 benchmark; it is not an original QASMBenchmarks circuit.}
\label{tab:osr-bqskit-four-qubit-qasm-benchmarks-counts}
\end{table*}

\begin{table*}[t]
\centering
\begin{tabular}{l|rrr|rrr}
\hline
Circuit & \multicolumn{3}{c|}{Squander time (min)} & \multicolumn{3}{c}{BQSKit time (min)} \\
 & A2A & Routing & Final & A2A & Routing & Final \\
\hline
\texttt{adder\_n4} & \textbf{0.12} & 0.65 & \textbf{0.09} & 63.37 & \textbf{0.23} & 49.69 \\
\texttt{bv\_n14} & \textbf{0.01} & 1.02 & \textbf{0.09} & 0.22 & \textbf{0.19} & 1.35 \\
\texttt{cat\_state\_n22} & -- & -- & \textbf{0.00} & -- & -- & 0.22 \\
\texttt{ghz\_state\_n23} & -- & -- & \textbf{0.00} & -- & -- & 0.26 \\
\texttt{multiply\_n13} & \textbf{1.12} & 2.34 & \textbf{5.14} & 39.11 & \textbf{0.45} & 69.80 \\
\texttt{tfim} & -- & -- & \textbf{5.29} & -- & -- & 366.97 \\
\texttt{bv\_n140} & \textbf{0.02} & 17.83 & \textbf{2.14} & 0.36 & \textbf{0.72} & 21.10 \\
\texttt{swap\_test\_n25} & \textbf{1.24} & 5.09 & \textbf{9.93} & 17.38 & \textbf{4.15} & 15.36 \\
\texttt{ising\_n10} & -- & -- & \textbf{1.80} & -- & -- & 66.07 \\
\texttt{grover\_19} & \textbf{1.52} & 8.86 & \textbf{6.68} & 320.21 & \textbf{2.17} & 89.68 \\
\texttt{ising\_n98} & -- & -- & \textbf{1.72} & -- & -- & 4.01 \\
\texttt{hhl\_n7} & \textbf{6.52} & 7.15 & \textbf{5.14} & 851.19 & \textbf{0.94} & 178.91 \\
\texttt{grover\_23} & \textbf{1.64} & 11.27 & \textbf{7.69} & 675.88 & \textbf{2.35} & 111.43 \\
\texttt{multiplier\_n15} & \textbf{3.78} & 12.82 & \textbf{18.14} & 626.61 & \textbf{1.23} & 291.27 \\
\texttt{heisenberg-16-20} & -- & -- & \textbf{4.41} & -- & -- & 71.58 \\
\texttt{grover\_33} & \textbf{2.01} & 17.85 & \textbf{9.84} & 294.82 & \textbf{3.56} & 91.16 \\
\texttt{grover\_45} & \textbf{2.69} & 26.63 & \textbf{7.21} & 359.53 & \textbf{4.41} & 199.25 \\
\texttt{grover\_61} & \textbf{3.99} & 62.47 & \textbf{9.15} & 570.96 & \textbf{6.34} & 113.37 \\
\hline
Total & \textbf{24.66} & 173.98 & \textbf{94.44} & 3819.65 & \textbf{26.74} & 1741.47 \\
\hline
\end{tabular}
\caption{QASMBenchmarks four-qubit partition stage runtimes for Squander and BQSKit at matched partition resynthesis infidelity tolerance, in minutes. Bold values are best within each paired stage. The tfim row is the QFAST/QSearch tfim-4-60 benchmark, it is not an original QASMBenchmarks circuit.}
\label{tab:osr-bqskit-four-qubit-qasm-benchmarks-times}
\end{table*}

For the four-qubit QASMBenchmarks (Tables~\ref{tab:osr-bqskit-four-qubit-qasm-benchmarks-counts} and~\ref{tab:osr-bqskit-four-qubit-qasm-benchmarks-times}), Squander's CNOT advantage is significant at all three stages, with the clearest per-circuit gains after routing: it uses fewer CNOTs in 11 of 12 Routed comparisons, with a median reduction of 62 CNOTs. The A2A reductions are smaller (7 wins and 5 ties, median reduction 3.5), while Squander also has fewer CNOTs in 12 of 18 Final comparisons, with ties in the other six (median reduction of 20.5 CNOTs). The A2A and Routed comparisons each include 12 circuit pairs, whereas Final includes 18 (the six additional circuits already satisfy the linear topology), so their totals describe different paired samples and should be interpreted stage by stage. Runtime shows a distinct trade-off. Squander is faster in every paired A2A and Final run, its median times are 1.58 versus 307.51 minutes and 5.14 versus 70.69 minutes, respectively, and the summed times favor Squander in both stages as well. Routing reverses this pattern: Squander has higher recorded times in all 12 pairs, with a higher median time (10.06 versus 1.70 minutes) and higher summed time (173.98 versus 26.74 minutes). Overall, four-qubit Squander provides a consistent CNOT benefit on these paired benchmarks and substantial A2A and Final runtime savings relative to BQSKit, but the routing stage remains slower. The table totals show 596 fewer Squander CNOTs in A2A and 1,170 fewer after routing across 12 paired circuits, plus 983 fewer at Final across 18 paired circuits.

\begin{table*}[t]
\centering
\begin{tabular}{lrr|rr|rr|rr}
\hline
Circuit & Qubits & Initial CNOTs & \multicolumn{2}{c|}{A2A} & \multicolumn{2}{c|}{Routed} & \multicolumn{2}{c}{Final} \\
 & & & CNOTs & Time (min) & CNOTs & Time (min) & CNOTs & Time (min) \\
\hline
\texttt{qft\_20} & 20 & 410 & 319 & 104.90 & 865 & 32.01 & 815 & 78.84 \\
\texttt{dnn\_n8} & 8 & 192 & 61 & 12.59 & 105 & 8.19 & 97 & 12.09 \\
\texttt{dnn\_n16} & 16 & 384 & 122 & 23.07 & 227 & 19.65 & 207 & 10.84 \\
\texttt{grover\_81} & 81 & 936 & 589 & 6.28 & 859 & 91.54 & 855 & 11.41 \\
\shortstack[l]{\texttt{quantum\_volume}\\\texttt{n40\_d20\_i0}} & 40 & 1200 & 1170 & 233.99 & 9264 & 226.66 & 7914 & 509.09 \\
\texttt{adder\_n64} & 64 & 455 & 266 & 2.83 & 361 & 20.00 & 328 & 8.09 \\
\hline
Total & -- & 3577 & 2527 & 383.67 & 11681 & 398.05 & 10216 & 630.37 \\
\hline
\end{tabular}
\caption{Completed QASMBenchmarks four-qubit partition Squander results for circuits without a completed BQSKit four-qubit result. Counts and stage runtimes are combined because no paired comparison is available.}
\label{tab:osr-four-qubit-unpaired-qasm-benchmarks}
\end{table*}
Comparing Squander across max partition sizes (3 and 4 qubits) on the QASMBenchmarks with results at both partition sizes (Tables~\ref{tab:osr-bqskit-three-qubit-qasm-benchmarks-counts},~\ref{tab:osr-bqskit-four-qubit-qasm-benchmarks-counts}, and~\ref{tab:osr-four-qubit-unpaired-qasm-benchmarks}) shows modest aggregate reductions at all three stages. For stage A2A the three- and four-qubit partition totals are 4,379 and 4,230 CNOT gates. For the stage Routed the same quantities are 14,424 and 14,384 CNOT gates, across the 18 circuits with recorded routing stages. For stage Final, the totals are 13,604 and 13,242 CNOT gates across all 24 completed circuits. Thus, four-qubit partitions lower all three aggregate stage counts, although the Routed difference is small. These aggregates do not imply that every circuit benefits. The four-qubit totals include Squander-only results presented in Table~\ref{tab:osr-four-qubit-unpaired-qasm-benchmarks} for six circuits without a completed BQSKit four-qubit result.

Across the completed IBM Eagle and QASMBenchmarks results (Tables~\ref{tab:osr-bqskit-three-qubit-ibm-eagle-counts},~\ref{tab:osr-bqskit-three-qubit-qasm-benchmarks-counts},~\ref{tab:osr-bqskit-four-qubit-ibm-eagle-counts},~\ref{tab:osr-four-qubit-unpaired-ibm-eagle},~\ref{tab:osr-bqskit-four-qubit-qasm-benchmarks-counts}, and~\ref{tab:osr-four-qubit-unpaired-qasm-benchmarks}), A2A resynthesis reduces CNOT counts relative to the initial circuits for both partition sizes. For the 53 circuits the initial total is 19,219 CNOTs. For stage A2A the total is 16,273 CNOTs with three-qubit partitions and 15,569 with four-qubit partitions, reductions of 15.3\% and 19.0\%, respectively, compared to the initial values. 
Mapping to the target topology then increases the counts substantially. Final optimization removes some of this routing overhead: on the same 53 circuits, it reduces Routed totals from 48,367 CNOTs to 44,870 CNOTs for three-qubit partitions (7.2\%), and from 51,574 to 45,440 CNOTs for four-qubit partitions (11.9\%). Although four-qubit partitions yield a slightly greater reduction from Routed to Final, the Final totals for both settings remain above the initial count due to the topology constraint.

\clearpage

Routing can substantially change CNOT counts, making the contribution of partition resynthesis difficult to distinguish in an end-to-end comparison. We therefore compare OSR and BQSKit one-to-one on the same partition unitaries, holding each target fixed. This isolates local synthesis quality and efficiency from differences in partition selection and routing.

To isolate the efficiency of local synthesis, Table~\ref{tab:local-osr-bqskit} compares matched three- and four-qubit targets from frozen benchmark-derived and random samples. The complete frozen cohort was rerun on 22 October 2026 using the current native build. Both methods receive the same unitary, coupling graph, unitary decomposition-infidelity threshold of $10^{-10}$, and a 30-second process watchdog, including startup, and a 4-GiB memory limit per solve. Each solve is pinned to one CPU, and the local OSR benchmark uses double precision. The table reports solver-only median times on jointly verified targets. Exported circuits are checked independently against the target and topology.
Quality comparisons include only jointly verified pairs, while completion rates are reported separately. The sample excludes compressed source archives larger than 8 MiB. This frozen corpus is a calibration benchmark rather than an independent held-out evaluation.
\begingroup
\AddToHookNext{env/tabular/before}{\setlength{\tabcolsep}{3pt}}
\begin{table*}[t]
\centering
\resizebox{\textwidth}{!}{%
\begin{tabular}{lrrrrrccc}
\hline
Context & Targets & Paired & OSR & BQSKit & OSR fail and & OSR & $\Delta$CX & Paired time \\
 & & & valid & valid & BQSKit valid & W/T/L & (95\% CI) & OSR/BQSKit (s) \\
\hline
3q/a2a & 147 & 147 & 147 & 147 & 0 & 87/60/0 & $-1.24\,[-1.46,-1.03]$ & $0.50/3.16$ \\ 
3q/random\_a2a & 32 & 32 & 32 & 32 & 0 & 26/6/0 & $-1.94\,[-2.50,-1.41]$ & $3.17/4.91$ \\ 
3q/random\_path & 32 & 32 & 32 & 32 & 0 & 28/4/0 & $-2.53\,[-3.19,-1.91]$ & $0.33/2.66$ \\ 
3q/routing\_catalog & 83 & 83 & 83 & 83 & 0 & 65/18/0 & $-1.33\,[-1.52,-1.10]$ & $0.24/2.25$ \\ 
3q/topology & 143 & 143 & 143 & 143 & 0 & 99/44/0 & $-1.29\,[-1.57,-1.04]$ & $0.02/2.06$ \\ 
4q/a2a & 81 & 32 & 80 & 32 & 0 & 25/7/0 & $-1.75\,[-2.43,-1.21]$ & $1.02/10.04$ \\ 
4q/random\_a2a & 32 & 15 & 21 & 15 & 0 & 14/1/0 & $-2.07\,[-3.13,-1.20]$ & $0.03/4.08$ \\ 
4q/random\_cycle & 32 & 16 & 25 & 16 & 0 & 15/1/0 & $-2.06\,[-3.25,-1.19]$ & $0.03/4.07$ \\ 
4q/random\_path & 32 & 21 & 30 & 21 & 0 & 21/0/0 & $-2.10\,[-2.62,-1.57]$ & $0.03/4.54$ \\ 
4q/random\_star & 32 & 20 & 31 & 20 & 0 & 19/1/0 & $-2.50\,[-3.60,-1.60]$ & $0.03/4.70$ \\ 
4q/topology & 126 & 76 & 117 & 76 & 0 & 67/9/0 & $-3.36\,[-3.88,-2.80]$ & $0.05/11.34$ \\ 
\hline
\end{tabular}%
}
\caption{Paired local synthesis of 3- and 4-qubit target unitaries of the partitions. Each verified exported circuit reproduces the same target up to global phase with decomposition infidelity at most $10^{-10}$ and obeys the stated coupling graph. Wins/ties/losses and mean CNOT differences use only jointly verified pairs. {For each paired target, $\Delta$CX is the OSR CNOT count minus the BQSKit count; the table reports its mean, giving every target equal weight. Negative values favor OSR. The 95\% confidence intervals are obtained from 2,000 samples of the original source circuits, drawn with replacement. All targets from a selected source circuit are included together, and the mean is recomputed over the included targets. Each independently generated random target forms its own group.} Median solver times use only jointly verified targets. "OSR fail and BQSKit valid" counts OSR timeouts or failures when BQSKit produces a verified circuit. {If either method was not run on a target, that target is excluded from the paired comparison. A method that was not run is not counted as a timeout or failure, and its absence does not give the other method a win.}}
\label{tab:local-osr-bqskit}
\end{table*}

\endgroup

Across every listed context, the mean CNOT difference favors OSR: each reported 95\% confidence interval for $\Delta$CX lies below zero, and OSR has no losses among jointly verified pairs, though some are ties. This advantage appears for both all-to-all and topology-constrained targets, and is largest in the four-qubit topology context ($\Delta$CX = $-3.36$). OSR also has a lower reported median solver time in every context. These quality comparisons are paired: three-qubit results are complete for both methods, whereas the four-qubit comparisons cover smaller jointly verified subsets. Completion rates should therefore be considered alongside paired quality.

{The source code and benchmark reproduction tools are available in the SQUANDER repository~\cite{SQUANDERRepository}. The end-to-end benchmark archive contains the input and optimized OpenQASM circuits, per-circuit configurations and source fingerprints, and stage-wise CNOT counts and runtimes for OSR and BQSKit at both partition sizes. Recorded failures and timeouts are retained alongside completed results. Compressed replay audits record the local rewrites and circuit-composition steps leading to each completed output.\footnote{{\color{black}Original benchmark archive: \url{https://github.com/rakytap/sequential-quantum-gate-decomposer/releases/tag/benchmark-results-20261028}. Refreshed result records and the evidence manifest are in \nolinkurl{benchmarks/partitioning/results_archive_20261111/} in the source repository.}}}

\section{Conclusion}

{
Our central contribution in this work is a change in how quantum-circuit optimization can be approached. Instead of committing to a single heuristic partition, we consider the full space of admissible subcircuit blocks  and select a globally best combination according to the resynthesis costs we compute. This makes global partition optimization practical at the circuit scales up to hundreds of qubits and thousands of quantum gates, with optimality defined over the enumerated candidates and their computed costs. We pair this broader search with an OSR-guided synthesis method that turns operator-Schmidt information from a bound into constructive guidance for discovering CNOT structures, and with a routing strategy that can compare alternative block covers using cached implementations and wire permutations. Together, these advances connect partitioning, synthesis, and routing, opening a path beyond one-shot heuristics toward more coordinated circuit optimization.}

The partitioning method enumerates admissible closed convex subcircuits, assigns their computed resynthesis costs, and selects an acyclic exact cover by ILP. Single-qubit-chain contraction and qubit-number-aware reachability reduce candidate-generation cost. Cycle-elimination constraints exclude covers that cannot be scheduled. Global optimality here means optimal partition selection over the enumerated feasible candidates and their computed costs, not a proof that the final circuit has the minimum possible CNOT count.

The OSR contribution uses operator-Schmidt spectra as constructive guidance for exploring CNOT structures, not merely as a posteriori descriptors. After Hilbert--Schmidt fitting of a candidate structure, all-cut dyadic rank bounds and $\kappa$-weighted residual scores prioritize topology-respecting extensions. A zero residual bound still requires a tighter fixed-structure synthesis check. This separation of fitting, structural guidance, and certification is the method evaluated here.


The stage-wise end-to-end tables show the outcomes of the complete SQUANDER and BQSKit workflows. The matched local-synthesis study provides the cleaner OSR comparison: across 772 fixed unitary decomposition targets, OSR verifies 741 outputs versus 617 for BQSKit, with 466 CNOT-count wins, 151 ties, and no losses on the 617 jointly verified pairs. 
{Across the 53 circuits, the initial total is 19,219 CNOT gates. OSR based A2A resynthesis reduces this to 16,273 with three-qubit partitions and to 15,569 with four-qubit partitions, meaning reductions of 15.3\% and 19.0\%, respectively. {\color{black}Routing to the target linear topology raises the totals} to 48,367 and 51,574 CNOTs. Final optimization then reduces the routed totals again to 44,870 (7.2\%) and 45,440 (11.9\%), respectively. Although four-qubit partitions yield a larger reduction from Routed to Final, both final totals remain above the initial count, highlighting the substantial effect of topology-aware routing on end-to-end circuit cost.}
{In paired comparisons with BQSKit, three-qubit SQUANDER saves 1,578 CNOTs (8.8\%) in A2A, 14,775 (23.4\%) after routing, and 14,839 (24.6\%) after Final optimization. For four-qubit partitions, the corresponding savings are 809 (19.0\%), 1,994 (23.1\%), and 1,729 (20.9\%).} Larger blocks expose additional synthesis opportunities, but the end-to-end benefit depends on routing as well. {{{\color{black}Further work will study routing objectives and larger synthesis windows.}}}

\begin{acknowledgements}
This research was supported by the Ministry of Culture and Innovation and the National Research, Development and Innovation Office through the Quantum Information National Laboratory of Hungary (Grant No.~2022-2.1.1-NL-2022-00004), financed by the National Research, Development and Innovation Fund, and by the Hungarian Scientific Research Fund (OTKA) under Grants No.~K134437. We also acknowledge support received through the EUREKA project AI4QT---Artificial Intelligence for Quantum Technologies (Grant No.~2020-1.2.3-EUREKA-2022-00029). We further acknowledge support from the Horizon Europe programme HORIZON-CL4-2022-QUANTUM-01-SGA through the OpenSuperQPlus100 project (Grant No.~101113946). We acknowledge the computational resources provided by the Wigner Scientific Computational Laboratory (WSCLAB), formerly the Wigner GPU Laboratory.
\end{acknowledgements}

\section*{AI assistance disclosure}
OpenAI GPT-5.5/5.6/6.0/6.1 Sol models, used through the Codex harness in medium and extra-high reasoning modes, assisted with method development, implementation, computational experiments, literature and citation checks, and manuscript preparation. The authors remain responsible for all scientific claims, computations, references, and text.

\appendix

\pagebreak
\section{Supplementary Algorithms and Results}

\subsection{Qubit Number-Aware Reachability and Convex Partition Enumeration}
\label{apx:qubit-number-aware-enum}

This appendix gives the pseudocode for the qubit number-aware reachability and convex partition enumeration routines used in Section~\ref{sec:ilp}. The main text describes the role of these routines in pruning the partition search space; here we provide the implementation-level details. Algorithm~\ref{alg:qubit_reach} restricts reachability to subgraphs whose cumulative qubit footprint remains within the budget $Q_{\max}$. Algorithm~\ref{alg:partition_enum} then uses this restricted reachability inside a depth-first enumeration of convex, qubit-feasible candidate partitions.

\begin{algorithm*}[t]
\caption{Qubit Number-Aware Reachability}
\label{alg:qubit_reach}
\KwIn{
\begin{tabular}{ll}
$X$ & initial set of gates, \\
$g, rg$ & forward and reverse adjacency of circuit DAG, \\
$R$ & set of admissible gates, \\
$\mathcal{Q}$ & mapping gate $\mapsto$ qubits, \\
$Q_{\max}$ & maximum allowed qubits per partition.
\end{tabular}}
\KwOut{Set of gates reachable without exceeding qubit bound.}

Initialize \texttt{level} $\gets X$, \texttt{visited} $\gets \emptyset$, \texttt{nextlevel} $\gets \emptyset$, and qubit-sets $Q_s$\;

\While{\texttt{level} $\neq \emptyset$}{
    \ForEach{$v \in$ \texttt{level}}{
        Let $R_v \gets rg(v) \cap R$ \tcc*{backward restriction check}
        \If{$R_v \subseteq$ \texttt{visited}}{
            \eIf{$v \in X$}{
                $Q_s[v] \gets Q$\;
            }{
                $Q_s[v] \gets \mathcal{Q}(v) \cup \bigcup_{u \in R_v} Q_s[u]$\;
            }
            \If{$|Q_s[v]| \leq Q_{\max}$}{
                \texttt{visited} $\gets$ \texttt{visited} $\cup \{v\}$\;
                \texttt{nextlevel} $\gets$ \texttt{nextlevel} $\cup (g(v) \cap R)$\;
            }
            \Else{
                delete $Q_s[v]$\;
            }
        }
    }
    Swap(\texttt{level}, \texttt{nextlevel}); clear(\texttt{nextlevel})\;
}
\Return set of vertices with defined $Q_s$\;
\end{algorithm*}

\begin{algorithm*}[t]
\caption{Qubit Number-Aware Convex Partition Enumeration}
\label{alg:partition_enum}
\KwIn{
\begin{tabular}{ll}
$G=(V,E)$ & circuit DAG, with topological order $\pi$, \\
$g, rg$ & forward and reverse adjacency, \\
$\texttt{reach}, \texttt{revreach}$ & precomputed closures, \\
$\mathcal{Q}(v)$ & qubit set of each gate, \\
$Q_{\max}$ & maximum qubits per partition.
\end{tabular}}
\KwOut{Set $\mathcal{P}$ of all convex qubit-feasible partitions.}

$\mathcal{P} \gets \emptyset$\;
\ForEach{vertex $t$ in topological order}{
  $X \gets \{t\},~Y \gets \{v:\pi(v)>\pi(t)\},~Q \gets \mathcal{Q}(t)$\;
  $A \gets \texttt{reach}[t] \cap Y,~~B \gets \texttt{revreach}[t] \cap Y$\;
  prune $Y, A$ using Alg.~\ref{alg:qubit_reach} with $(X,g,rg,A,Q)$\;
  stack $\gets \{(X,Y,A,B,Q)\}$\;

  \While{stack not empty}{
    Pop $(X,Y,A,B,Q)$\;
    \If{$A \neq \emptyset$}{select $v \in A$ and set $R \gets (A \cap \texttt{revreach}[v]) \cup \{v\}$}
    \ElseIf{$B \neq \emptyset$}{select $v \in B$ and set $R \gets (B \cap \texttt{reach}[v]) \cup \{v\}$}
    \Else{$\mathcal{P} \gets \mathcal{P} \cup \{X\}$; continue}
    push $(X, Y \setminus \{v\}, {A \setminus \{v\}}, {B \setminus \{v\}}, Q)$ to stack\;
    $Y' \gets Y \setminus R,~~X' \gets X \cup R,~~Q' \gets Q \cup \bigcup_{u \in R} \mathcal{Q}(u)$\;
    \If{$|Q'| \leq Q_{\max}$}{
       $A' \gets (A \setminus R) \cup \bigcup_{u \in R} (\texttt{reach}[u]\cap Y')$\;
       $B' \gets (B \setminus R) \cup \bigcup_{u \in R} (\texttt{revreach}[u]\cap Y')$\;
       prune $Y', A'$ with Alg.~\ref{alg:qubit_reach} $(X',g,rg,A',Q')$\;
       prune $Y', B'$ with Alg.~\ref{alg:qubit_reach} $(X',rg,g,B',Q')$\;
       push $(X',Y',A',B',Q')$ to stack\;
    }
  }
}
\Return $\mathcal{P}$\;
\end{algorithm*}
\paragraph*{Preprocessing.} 
The forward and backward reachability closures for sets $A$ and $B$ are computed using Nuutila's transitive-closure algorithm~\cite{NUUTILA1994207}, applied to the graph and its reverse:
\begin{equation}
\texttt{reach}[u] = \{ v \mid u \preceq v \}, \quad 
\texttt{revreach}[u] = \{ v \mid v \preceq u \}.
\end{equation}
These closures are then indexed by a topological order $\pi : V \to \{1,\dots,n\}$ of the DAG.
For each topologically earliest seed gate $t \in V$, the algorithm initiates a convex set $X = \{t\}$ with remaining candidate vertices $Y = \{ v : \pi(v) > \pi(t)\}$ and qubit footprint $\mathcal{Q}(t)$. The two frontier sets $A,B \subseteq Y$ are always maintained relative to the current partition $X$.
The search proceeds in a depth-first manner with a stack of states. At each step, one candidate vertex $v \in A \cup B$ is selected, and a region $R$ (all vertices forced in by convexity) is added to the current partition $X$. {\color{black}If the enlarged partition’s qubit set remains within $Q_{\max}$, the frontier sets are updated and pruned by the qubit-number-aware BFS reachability routine (Algorithm~\ref{alg:qubit_reach}); otherwise, the inclusion branch is discarded. When both frontiers are empty, $X$ is recorded as a partition. The process backtracks until all convex partitions rooted at $t$ have been enumerated.}

Using the outlined approach, each partition is discovered once, and each recursive expansion checks only admissible $Q_{\max}$-bounded frontiers. The worst-case complexity is exponential in $|V|$ (since the number of convex partitions can be exponential), but per-step overhead is polynomial, with $O(|E|)$ for reachability updates and $O(Q_{\max})$ for qubit-set unions. In practice, the qubit bound $Q_{\max}$ is small, so the algorithm scales to circuits with hundreds or thousands of gates.

\subsection{Algorithm for Contracting Single-Qubit Chains}\label{apx:contract}

In this appendix, we describe the procedure used to contract single-qubit chains within a quantum circuit. The algorithm operates on a directed acyclic graph (DAG) representation of the circuit, where vertices correspond to gates, and edges represent data dependencies. For each qubit, consecutive single-qubit gates are merged into chains to reduce the effective graph size while preserving circuit semantics.

\begin{algorithm*}[t]
\caption{Contracting Single-Qubit Chains}
\label{alg:contract_single_qubit_chains}
\KwIn{
\begin{tabular}{ll}
$G$ & forward adjacency map of the DAG (successors of each gate), \\
$G^{-1}$ & reverse adjacency map of the DAG (predecessors of each gate), \\
$Q(g)$ & mapping from each gate $g$ to the set of qubits it acts on, \\
$\mathcal{T}$ & a valid topological ordering of all gates.
\end{tabular}
}
\KwOut{
\begin{tabular}{ll}
$G'$ & contracted forward adjacency map, \\
$G'^{-1}$ & contracted reverse adjacency map, \\
$\mathcal{T}'$ & updated topological ordering excluding contracted gates, \\
$\mathcal{C}$ & dictionary mapping each representative gate to its contracted single-qubit chain.
\end{tabular}
}

\BlankLine
$\mathcal{S} \gets \{ g \in V(G) : |Q(g)| = 1 \}$ \tcc*{Set of single-qubit gates}
$\mathcal{C} \gets \emptyset$ \tcc*{Initialize contracted chains}
\ForEach{$g \in \mathcal{T}$}{
    \If{$g \in \mathcal{S}$}{
        \If{$G^{-1}(g) \neq \emptyset$}{
            let $p \gets$ an arbitrary predecessor of $g$ \;
            $\mathcal{C}[g] \gets \mathcal{C}[p]$ if $p \in \mathcal{C}$ else $\emptyset$\;
        }
        \Else{
            $\mathcal{C}[g] \gets \emptyset$\;
        }
        append $g$ to $\mathcal{C}[g]$\;
        
        \If{$G^{-1}(g) \neq \emptyset$}{
            let $p \gets$ an arbitrary predecessor of $g$ \;
            $G(p) \gets (G(p) \setminus \{g\}) \cup G(g)$\;
        }
        \If{$G(g) \neq \emptyset$}{
            let $s \gets$ an arbitrary successor of $g$ \;
            $G^{-1}(s) \gets (G^{-1}(s) \setminus \{g\}) \cup G^{-1}(g)$\;
        }
        remove $g$ from $G$ and $G^{-1}$\;
    }
}
$\mathcal{T}' \gets \mathcal{T} \setminus \mathcal{S}$\;
\Return $(G, G^{-1}, \mathcal{T}', \mathcal{C})$\;
\end{algorithm*}

\noindent
The algorithm proceeds as follows:
\begin{enumerate}
    \item Identify all single-qubit gates.
    \item Traverse the circuit in topological order, extending single-qubit chains whenever possible.
    \item For each gate in a chain, update adjacency relations by bypassing the contracted node, ensuring that dependency edges are preserved.
    \item Remove contracted nodes from the adjacency maps and the topological order.
\end{enumerate}
This contraction reduces the number of nodes considered during later partitioning stages, while a dictionary map (labeled by $\mathcal{C}$ in Alg.~\ref{alg:contract_single_qubit_chains}) maintains explicit records of the merged single-qubit gates to allow full reconstruction of the original circuit.

\paragraph*{Complexity Analysis}
Let $|V|$ denote the number of gates and $|E|$ the number of dependency edges in the circuit DAG. Identifying single-qubit gates requires $O(|V|)$ time. The traversal in topological order visits each gate once and processes each adjacency update in $O(1)$ expected time using hash-based adjacency maps. Since each edge is considered at most once when bypassing contracted gates, the total update cost is $O(|E|)$. Therefore, the total time complexity is
\[
O(|V| + |E|).
\]
The space complexity is also linear: adjacency maps store $O(|V| + |E|)$ entries, and the dictionary $\mathcal{C}$ introduces only an additional $O(|V|)$ storage overhead. Hence, the algorithm is asymptotically optimal with respect to both time and space for this task.

\subsection{Algorithm for Recombining Single-Qubit Chains}\label{apx:recomb}

After circuit partitioning has been performed on the contracted DAG, we must restore the original single-qubit gates to ensure correctness of the final partitioning. The following algorithm reinserts contracted single-qubit chains into the partition sets, maintaining consistency with gate dependencies.

\begin{algorithm*}[t]
\caption{Recombining Single-Qubit Chains}
\label{alg:recombine_single_qubit_chains}
\KwIn{
\begin{tabular}{ll}
$G$ & forward adjacency map of the contracted DAG, \\
$G^{-1}$ & reverse adjacency map of the contracted DAG, \\
$\mathcal{C}$ & dictionary of single-qubit chains (as returned by Algorithm~\ref{alg:contract_single_qubit_chains}), \\
$\mathcal{L}$ & a partition of the contracted circuit gates into disjoint sets.
\end{tabular}
}
\KwOut{
$\mathcal{L}'$, a partition of the full circuit including recombined single-qubit chains.
}

\BlankLine
$\mathcal{L} \gets \{ \text{set}(P) : P \in \mathcal{L}\}$ \tcc*{Convert partitions to mutable sets}
initialize map $M : g \mapsto P$ such that $g \in P$ for each gate $g$ and partition $P \in \mathcal{L}$\;

\ForEach{chain $C \in \text{unique}(\mathcal{C})$}{
    let $g_{\mathrm{head}} \gets C[0]$ and $g_{\mathrm{tail}} \gets C[-1]$\;
    
    \If{$G^{-1}(g_{\mathrm{head}}) \neq \emptyset$}{
        let $p \gets$ an arbitrary predecessor of $g_{\mathrm{head}}$\;
        $M(p) \gets M(p) \cup C$\;
    }
    \ElseIf{$G(g_{\mathrm{tail}}) \neq \emptyset$}{
        let $s \gets$ an arbitrary successor of $g_{\mathrm{tail}}$\;
        $M(s) \gets M(s) \cup C$\;
    }
    \Else{
        $\mathcal{L} \gets \mathcal{L} \cup \{ C \}$ \tcc*{Chain is isolated, forms a new partition}
    }
}
\Return $\mathcal{L}$\;
\end{algorithm*}

\noindent
The procedure works by traversing each contracted chain $C$ and merging it back into the partition containing one of its neighboring gates:
\begin{enumerate}
    \item If the chain has a predecessor in $G^{-1}$, it is merged into the same partition as that predecessor.
    \item Otherwise, if the chain has a successor in $G$, it is merged into that partition.
    \item If the chain is isolated (no predecessors or successors), it is appended as its own partition.
\end{enumerate}

\paragraph*{Complexity Analysis}
Let $|V|$ denote the number of contracted gates, $|E|$ the number of edges, and $|\mathcal{C}|$ the number of single-qubit chains. Constructing the mapping $M$ requires $O(|V|)$ time. Each chain is processed exactly once, with constant-time set merges and membership lookups using hash tables. The total number of insertions across all chains is $O\!\left(\sum_{C \in \mathcal{C}} |C|\right) = O(|V|)$, since each gate belongs to at most one chain. Therefore, the overall time complexity is
\[
O(|V| + |\mathcal{C}|).
\]
The space complexity is $O(|V| + |\mathcal{C}|)$, dominated by the partition map $M$ and storage of chain memberships.

\bibliographystyle{quantum}
\bibliography{biblio}

\end{document}